\documentclass[aps,twocolumn,showpacs,preprintnumbers,amsmath,amssymb,superscriptaddress]{revtex4}
\usepackage{xcolor}
\usepackage{graphicx}%
\usepackage{dcolumn}%
\usepackage{bm}%
\usepackage{color}
 \usepackage{subfigure}

\begin{document}

\title{Phase solitons and domain dynamics in an optically injected semiconductor laser.}%

\author{F. Gustave}
\affiliation{Universit\'e de Nice CNRS, Institut Non Lin\'eaire de Nice, 1361 route des lucioles 06560 Valbonne, France}
\author{L. Columbo}
\affiliation{Dipartimento Interateneo di Fisica, Universit$\grave{a}$ degli Studi e Politecnico di Bari, Via Amendola 173, 70126 Bari, Italy}%
\author{G. Tissoni}
\affiliation{Universit\'e de Nice CNRS, Institut Non Lin\'eaire de Nice, 1361 route des lucioles 06560 Valbonne, France}
\author{M. Brambilla}
\affiliation{Dipartimento Interateneo di Fisica, Universit$\grave{a}$ degli Studi e Politecnico di Bari, Via Amendola 173, 70126 Bari, Italy}%
\author{F. Prati}
\affiliation{Dipartimento di Scienza e Alta Tecnologia, Universit$\grave{a}$ dell'Insubria, Via Valleggio 11, 22100 Como, Italy}%
\author{S. Barland}
\email{stephane.barland@inln.cnrs.fr}
\affiliation{Universit\'e de Nice CNRS, Institut Non Lin\'eaire de Nice, 1361 route des lucioles 06560 Valbonne, France}

\date{\today}%

\begin{abstract}

	We analyze experimentally and theoretically the spatio-temporal dynamics of a highly multimode semiconductor laser with coherent optical injection. Due to the particular geometry of the device (a 1~m long ring cavity), the multimode dynamics can be resolved in real time and we observe stable chiral solitons and domain dynamics. The experiment is analyzed in the framework of a set of effective semiconductor Maxwell-Bloch equations. We analyze the stability of stationary solutions and simulate both the complete model and a reduced rate equation model. This allows us to predict domain shrinking and the stability of only one chiral charge that we ascribe to the finite active medium response time.

\end{abstract}

\pacs{42.65.Tg, 42.65.Sf}%
\maketitle

\section{Introduction}

Optical systems are often used as an experimental test bench for the analysis of complex dynamical phenomena. In the context of laser physics, many such studies have been devoted to the laser with injected signal \cite{Wieczorek2005,lugiato2015nonlinear}. This interest can be attributed to the fact that the forcing term breaks the phase symmetry of the laser system. Doing so, it increases by one the number of dimensions of the phase space \cite{arnold1989mathematical}, and therefore brings the very common singlemode class-B lasers (semiconductor, CO$_2$, most solid-state) from bidimensional to three-dimensional, thus allowing chaotic dynamics. Because of their relative experimental ease of use and theoretical convenience, lasers with injected signal are therefore a widely explored topic whose study remains very lively on specific topics like optical excitability \cite{kelleher2009excitable,vaudel2008synchronization,Kelleher2010,turconi2013control}. However, most works to date have remained limited to single mode dynamics  with comparatively few works focussed on either in transverse \cite{coulletexcitwaves} or longitudinal \cite{Castelli1994,PhysRevA.32.1588,315223} spatiotemporal dynamics. Yet, multimode or spatially extended lasers with coherent forcing can be an extraordinary tool to explore synchronization and dynamics of oscillatory media with forcing, a conceptually simple yet very rich dynamical context \cite{PhysRevLett.56.724,coullet1992strong,chate199917}.

Here, following the recent observation of phase solitons hosting a chiral charge and deeply related to excitable dynamics \cite{PhysRevLett.115.043902}, we describe experimental observations of the spatio-temporal behavior of a strongly multimode semiconductor ring laser with optical forcing. We evidence plane-wave and modulational instabilities and show the propagation of fronts. In order to explain these new experimental findings we derive a set of effective semiconductor Maxwell-Bloch equations and analyze the stability of injection locked solutions with an exact and an approximated approach. The model is further reduced to a set of rate-equations which reproduce well the basic features of the relevant dynamical regimes and of the phase solitons, at difference from the Ginzburg-Landau derived in \cite{PhysRevLett.115.043902} which assumes instantaneous gain. \\

In section \ref{sec:experiment} we present the experimental device (\ref{subsec:setup}) followed by spectral analysis of the dynamics (\ref{subsec:spectral}). We then describe different instabilities (\ref{subsec:instabilities}) and analyze the propagation of solitons and fronts (\ref{subsec:solitons}). The theoretical analysis is reported in section \ref{sec:theory}, where we first derive a set of effective semiconductor Maxwell-Bloch equations (\ref{subsec:mbe}) and their reduction to rate-equations (\ref{subsec:rate}). We then analyze the injection locked solution and its stability (\ref{subsec:stability}) and describe the results of numerical simulations (\ref{subsec:simulations}). Finally, we present our conclusions in section \ref{sec:conclusions}

\section{Experiment}
\label{sec:experiment}
\subsection{Experimental setup}
\label{subsec:setup}
\begin{figure}[t]
\includegraphics[width=0.5\textwidth]{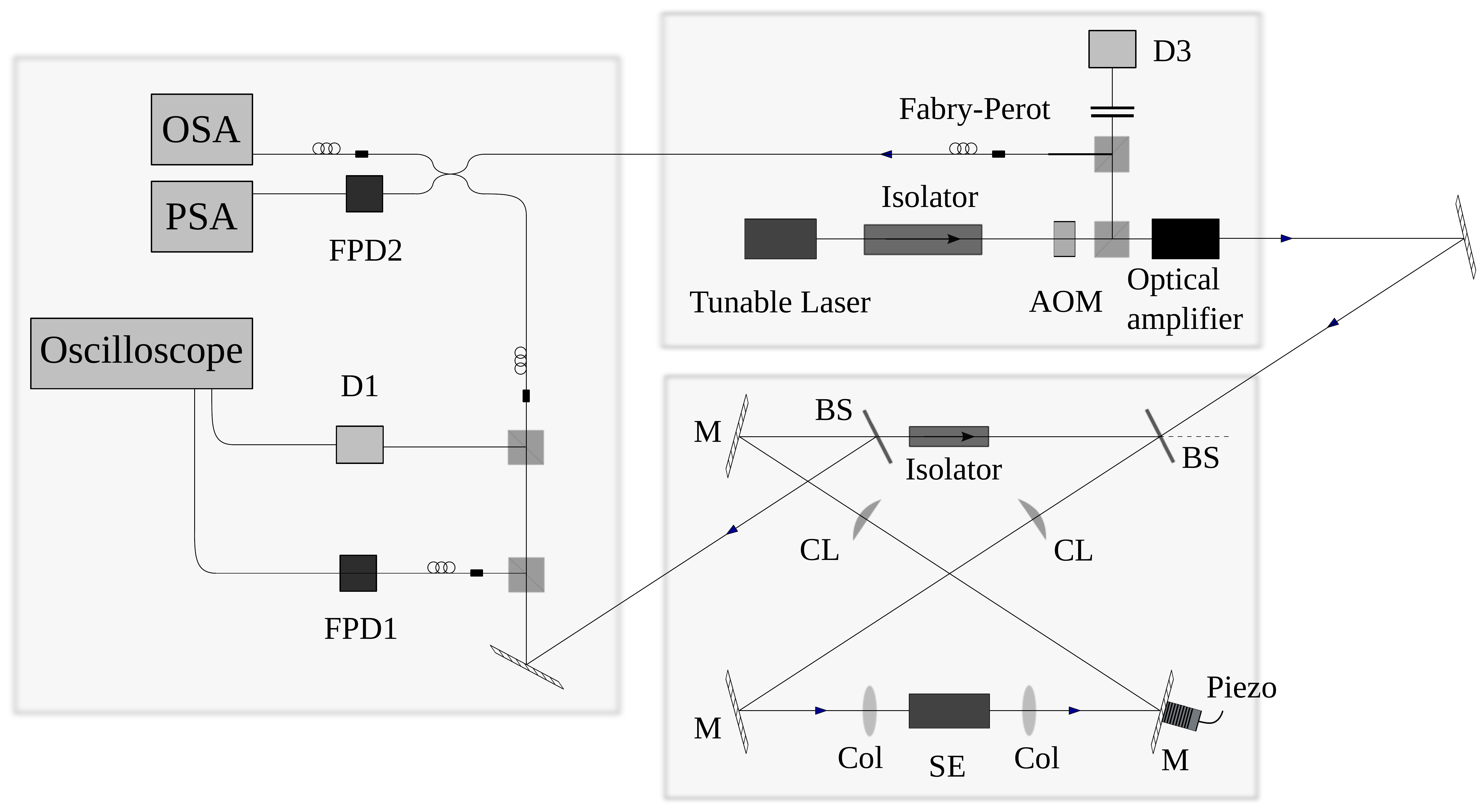}
\caption{Left: Scheme of the experimental setup. The semiconductor element SE is enclosed in a unidirectional ring cavity, which includes two beam splitters on of which is used for signal pick-up and the other one for external forcing. Cylindrical lenses CL are used to compensate for the very strong astigmatism of the beam caused by the aperture of the semiconductor element. FPD1-2 are high bandwidth (9~GHz) photodetectors used for time series (oscilloscope) and power spectrum (PSA) measurements. Detector D1 is used to monitor slow dynamics (micro to millisecond) associated to thermal effects. The optical spectrum analyzer is used for coarse tuning of the forcing beam close to the ring laser frequency. The tunable master laser spectrum is monitored by a Fabry-Perot interferometer and part of the forcing beam is sent to a fiber coupler for beat note or phase measurements. \label{fig:setup}}
\end{figure}

The experimental setup is shown in Fig. \ref{fig:setup}. It consists of a highly multimode semiconductor ring laser with external forcing. The ring laser is built by enclosing a 4~mm long, 980~nm, semiconductor optical amplifier inside a ring cavity. The direction of the junction is horizontal (\textit{ie} the largest transverse direction of the gain stripe is vertical). Due to the geometry of the gain stripe, the output beam of the optical amplifier is very astigmatic. 

At difference from standard operation of optical amplifiers, the present amplifier being operated in a laser with injected signal configuration requires very careful stabilization of temperature in order to keep the laser wavelength emission constant with respect to the external forcing. This is here achieved by two Peltier elements operating in parallel and with separate heat sinks. One of the Peltier elements provides the bulk of heat removal and is operated at constant current, while the other one is operated in a standard proportional-integral-derivative scheme for high-sensitivity active stabilization. The highly diverging direction is collimated on each side with high numerical aperture spherical collimators and the remaining divergence along the other direction (vertical) is then compensated for by a set of cylindrical lenses. The ring cavity is eight-shaped because it allows for lower incidence angle of the beam on the cavity mirrors. The laser is kept unidirectional by an optical isolator placed at the opposite side of the cavity with respect to the gain medium. The former, combined with the cylindrical lenses, has the additional effect of filtering out unwanted transverse effects \cite{Tierno:12} by constraining emission on a single transverse mode. The ring cavity itself is constituted by three high reflectivity mirrors (R$>$99\%) and one 90\% reflectivity beamsplitter used as an input for the forcing beam. One of the mirrors (bottom right in Fig. \ref{fig:setup}) is mounted on a piezoelectric actuator which allows subwavelength tuning of the cavity length. An additional 10\% reflectivity outcoupling beam splitter is inserted in the cavity for the detection path since the other available output (at the 90\% beamsplitter) is unusable due to very strong direct reflection of the forcing beam which brings the detectors to saturation. The roundtrip time is about 3.6~ns and the measured field lifetime is of the order of 10ns. This means a very low finesse (less than 10) of the cavity as compared to the value which could be estimated from the beam splitter and mirror reflectivities, which we attribute to poor mode matching (somewhat unavoidable with strongly astigmatic beams). 

The output of the laser is measured in the direction of the injection beam. The output beam is split in several parts, used for low bandwidth (but high sensitivity) emitted power measurement, optical spectrum measurement and high bandwidth (9~GHz) time-series or power spectrum measurement. The first two detection paths are used to optimize the ring laser alignment by minimizing standalone lasing threshold and to ensure proper tuning of the injection beam with respect to the ring laser. The emitted power in the operating regime (about $1.1$ times the threshold current) is of the order of 10~mW.

The injection beam is provided by a grating tunable external cavity semiconductor laser followed by an optical amplifier providing about 200~mW optical power. Although no emission is expected from the ring laser back into the master laser since the ring laser is constrained to be unidirectional, an optical isolator protects the master laser from eventual spurious backreflections.

\subsection{Spectral analysis}
\label{subsec:spectral}

\begin{figure}[t]
\center
\includegraphics[width=0.5\textwidth]{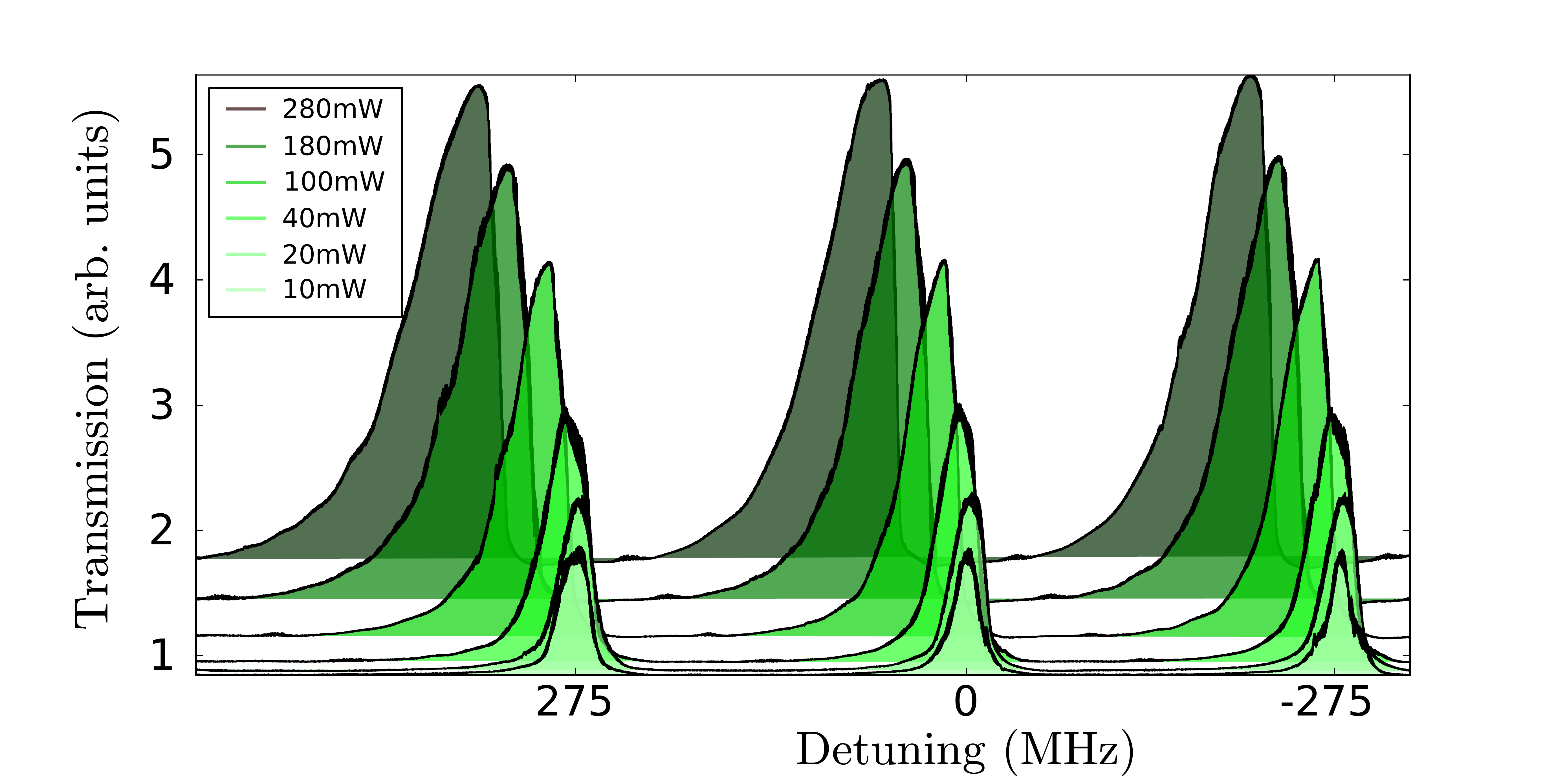}
\caption{Transmission of the ring cavity as a function of the detuning between the master laser and the resonances of the ring cavity, obtained for different values of injected power.}
\label{fig:transmission}
\end{figure}

The two most important control parameters of the experiment are injected power and detuning between master and slave lasers. We show  in Fig. \ref{fig:transmission} the transmission of the ring cavity when biased below standalone emission threshold depending on the detuning between the ring cavity and the injection beam $\Delta=\nu-\nu_c$, for several values of injected power. This figure has been obtained by scanning the voltage on the piezoelectric actuator and measuring the time averaged output power of the ring cavity. At low injected power the transmission curve shows three symmetric maxima. Each of these peaks correspond to the resonances of a linear optical cavity and are separated by 275~MHz as expected due to the length of the ring. Increasing the injected beam power the cavity resonances are shifted towards the red (up to about 60MHz)  and become asymmetric. While the left part of the resonance remains essentially unaltered, the right part becomes steeper and finally essentially vertical for increasing values of the injected power. We underline that these curves were obtained with a low-bandwidth detector, which averages out any spatial or temporal dynamics. In fact, time trace observations for static value of the voltage corresponding to the vertical edge of the resonance show very slow square wave oscillations due to thermal effects, analyzed in the context of slow-fast excitability in \cite{prethermal} and described as regenerative oscillations in \cite{spinelli2002thermal,tissoni2002spatio}. In \cite{prethermal,spinelli2002thermal,tissoni2002spatio}, these oscillations have been attributed to temperature dynamics, which slowly destabilizes both states of an optical bistability loop \footnote{Note that the temperature stabilization is as good as it can be: the dynamics results from the unavoidable relaxation of the active medium temperature towards the temperature of the substrate, which is the only one which can be controlled.}. In these condtions, no dynamics faster than the round-trip time has been observed.

\begin{figure}[t]
\center
\includegraphics[width=0.45\textwidth]{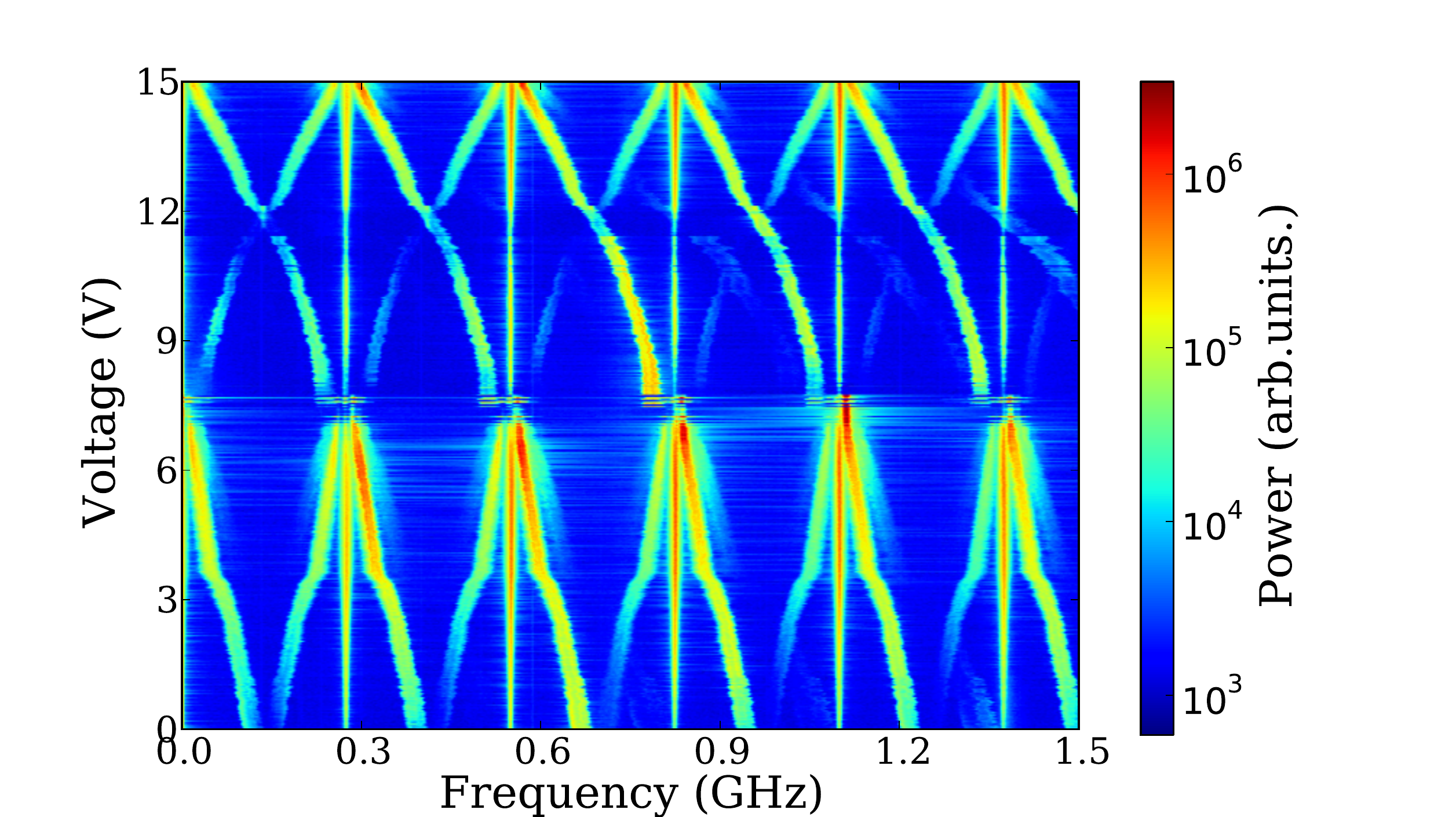}
\caption{Evolution of the power spectrum computed on time traces when the detuning is changed by applying voltage to the piezo. The dominant parabolic shapes correspond to the fact that the displacement of the piezoelectric component is not strictly linear with voltage.}
\label{fig:powerspectra}
\end{figure}

On the contrary spatial dynamics, of course associated with multimode instabilities in the context of optical resonators, can be readily observed when the bias current of the ring laser is brought above the lasing emission threshold. Even in absence of optical injection, the laser emission consists of several longitudinal modes. In presence of optical injection, many frequencies therefore show up in the power spectrum as plotted in Fig. \ref{fig:powerspectra}. The vertical lines correspond to the beat notes between successive longitudinal modes of the ring laser. Due to the slightly nonlinear response of the piezo element on the applied voltage, the actual detuning between the forcing and the laser modes varies quadratically when the voltage is varied. This translates in parabolic shapes for the beat note between forcing and the laser when the forcing is far from resonant conditions (voltage close to 0 or 10). Closer to resonant condition (between 3 and 9~V) the frequencies are pulled and the beat notes do not follow this parabolic shape. The actual detuning is varied in almost two free spectral ranges over the 0-15~V scan. In this particular case, the strongest frequency is close to 800~MHz, which indicates that the forcing frequency is detuned about three free spectral ranges from the dominant laser cavity mode. Nevertheless, many other longitudinal modes are also involved and, depending on specific conditions, the full width of the spectrum can reach about 10~GHz.

\subsection{Instabilities of the homogeneous solution}

\label{subsec:instabilities}

Time traces monitored by the high bandwidth detector contains all the information that can be extracted from a pure intensity measurement but actually this is not the most convenient way to visualize the data. In fact, most of the traces are exceedingly complex and the phenomena taking place can only be grasped by observing in a specific comoving reference frame. This is obtained by acquiring very long real time measurements (from 10 to 200 million points) and splitting this array of points in many segments of length equal to one cavity roundtrip time. These segments are then stacked on top of each other, which provide a space-time representation of the dynamics, the horizontal axis being equivalent to space inside the cavity while the vertical dimension describes the evolution in units of roundtrips. Many spatio-temporal regimes can be straightforwardly analyzed in this representation whereas some of them would be very difficult to grasp in a purely temporal representation. 

\begin{figure}[t]
\center
\subfigure[]{\includegraphics[width=0.235\textwidth]{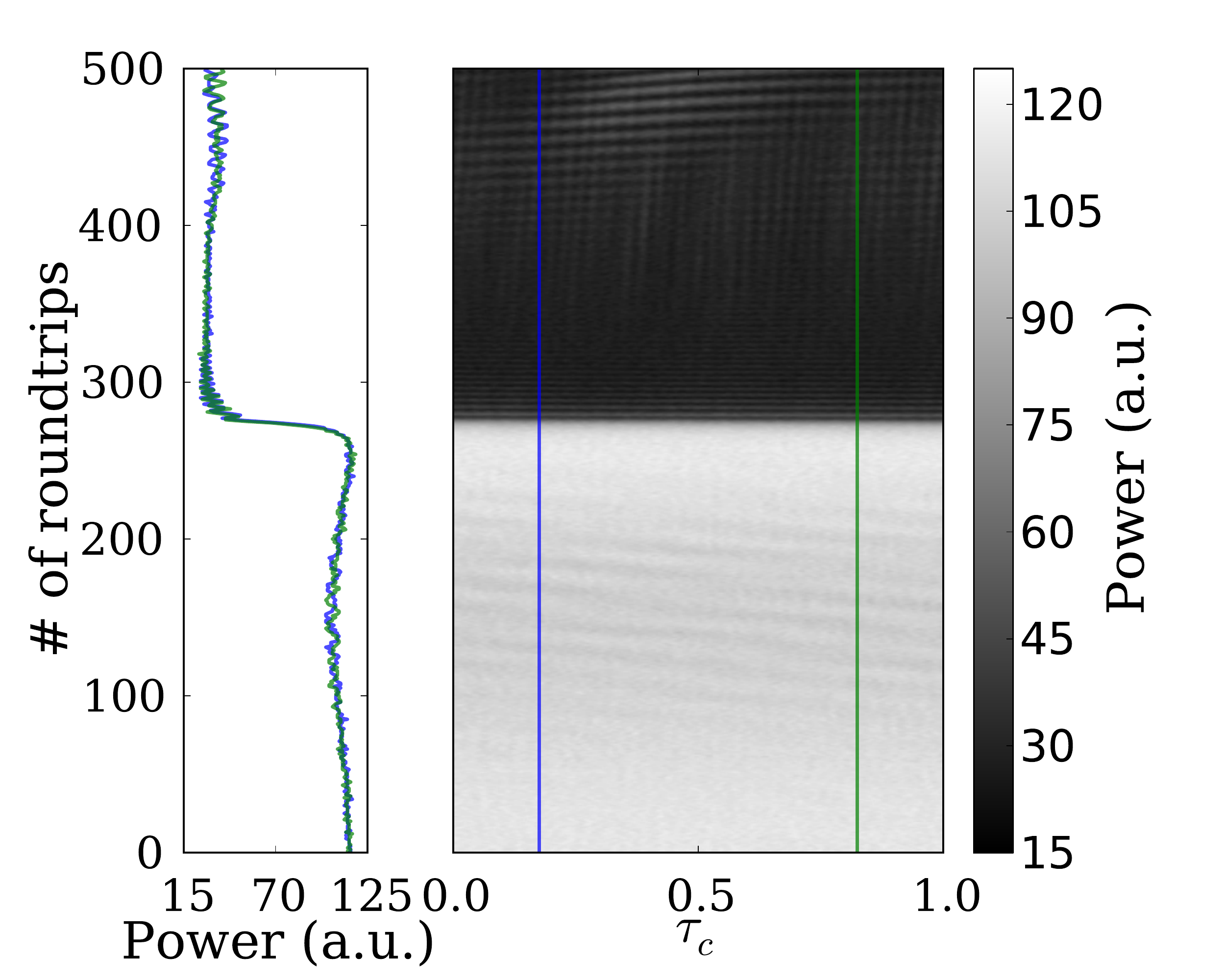}}
\subfigure[]{\includegraphics[width=0.22\textwidth]{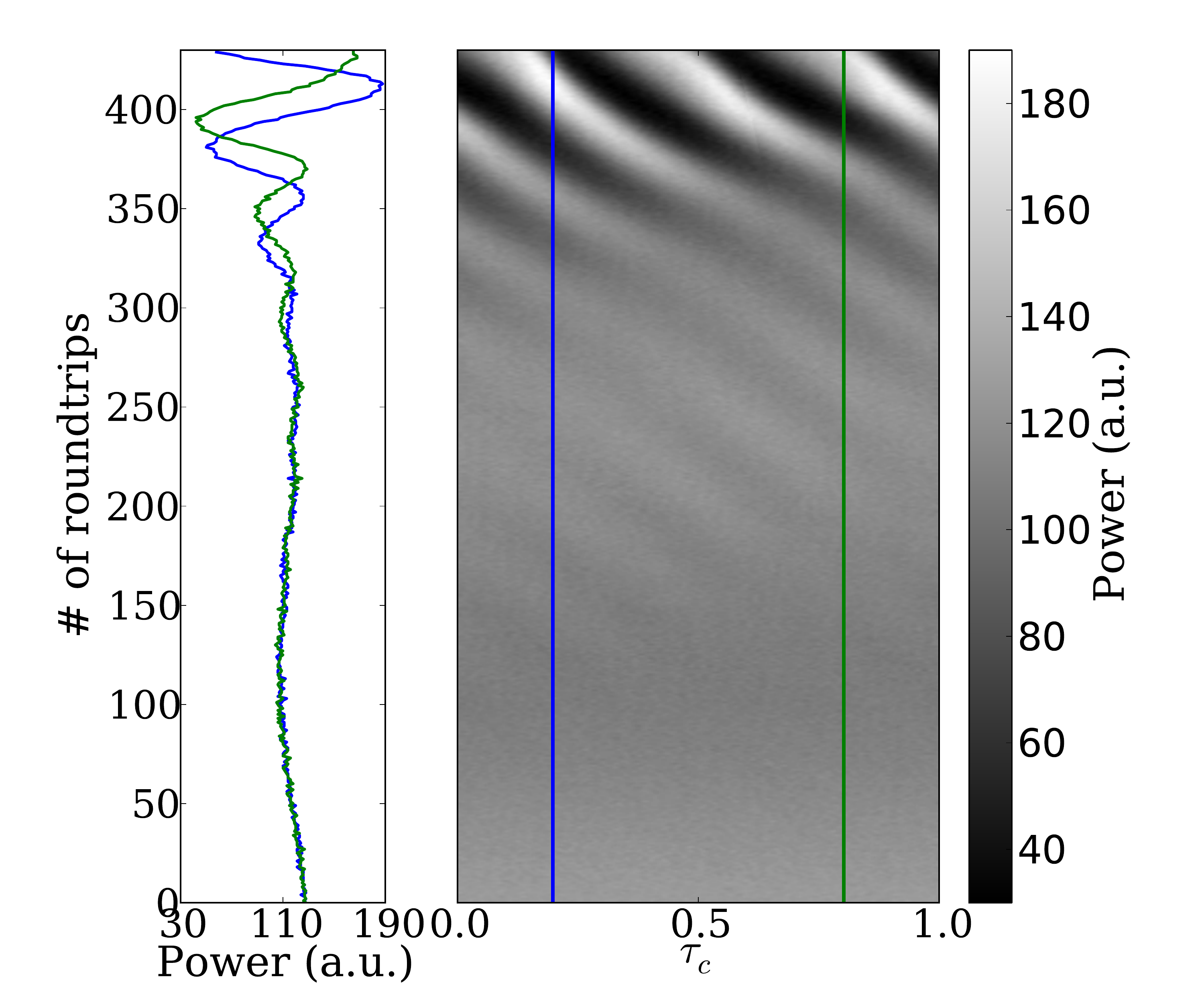}}
\caption{Different instabilities can develop along the propagation direction. (a) Plane wave instability, (b).}
\label{pl_instab}
\end{figure}

In Fig. \ref{pl_instab} (left panel) we show an example of a plane wave instability. Here the whole spatial extension of the system changes state at about roundtrip 260, switching from a high power to a low power state, both of them locked at the frequency of the external forcing. The two vertical time traces on the left correspond to the evolution of two different points of space (marked by the vertical lines in the space-time diagram), which appear to be very well synchronized. Also the small amplitude oscillations which appear between roundtrips 400 and 500 take place essentially along the whole spatial extension of the system, indicating a fundamentally single mode behavior. On the contrary, on the right panel of Figure \ref{pl_instab}, we show the growth of an instability which involves several longitudinal modes. In this case, starting from a stationary homogeneous solution a spatially periodic pattern develops and drifts.

\subsection{Solitons and fronts}
\label{subsec:solitons}
\begin{figure}[t]
\center
\includegraphics[width=0.5\textwidth]{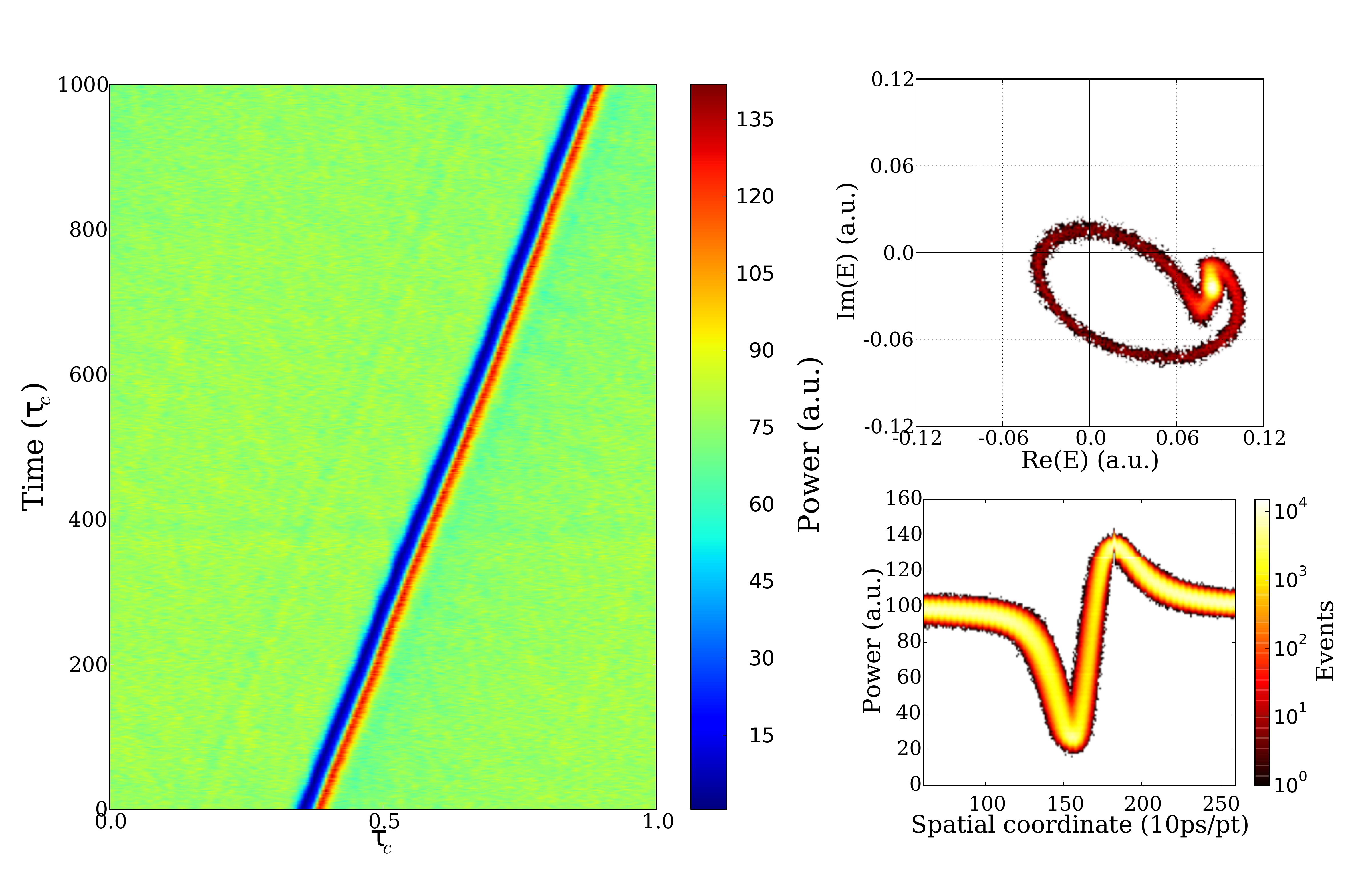}
\caption{Robustness of phase solitons. Left frame: a localized wave packet of 200~ps duration propagates at constant speed over 1000 roundtrips. It consists of a $2\pi$ phase rotation (top right frame), embedded in a uniformly phase locked background. Bottom right: color coded bidimensional histogram of soliton temporal profile (close to 1~ns duration), which shows remarkably constant shape over $4\times10^4$ roundtrips. Note the logarithmic color scale.}
\label{solshape}
\end{figure}

In Fig. \ref{solshape} (left frame), the laser beam intensity is uniform except for a narrow perturbation at about 0.35 space units. The width of this  pulse depends in a non trivial way on the detuning between the forcing and the closest laser mode but pulses as narrow as 200~ps have been observed. Note that this breadth amounts to about $1/18$ of the roundtrip time, as it will be commented further on. In this regime, the whole system is locked to the external forcing, except for the pulse, which also consists of a $2\pi$ relative phase rotation of the slave laser field with respect to the injected field (top right frame). In order to get an indication for the robustness of these wave packets, we have acquired time traces corresponding to 40~km propagation (\textit{i.e.} $4\times10^4$~roundtrips, which corresponds to the maximal record length of our oscilloscope). Then we have superimposed the corresponding $4\times10^4$ observations of the pulse and superimposed them accurately (effectively cancelling any jitter) on a color-coded bidimensional histogram mimicking analogue oscilloscope persistence (bottom-right frame). The very narrow resulting distribution indicates that these wavepackets are extremely robust, most of the dispersion of the curves actually resulting from electrical noise in the detector and oscilloscope. Since several of these non dispersing wave packets can coexist independently of each other, they have been analyzed as phase solitons in \cite{PhysRevLett.115.043902}.

\begin{figure}[t]
\center
\includegraphics[width=0.4\textwidth]{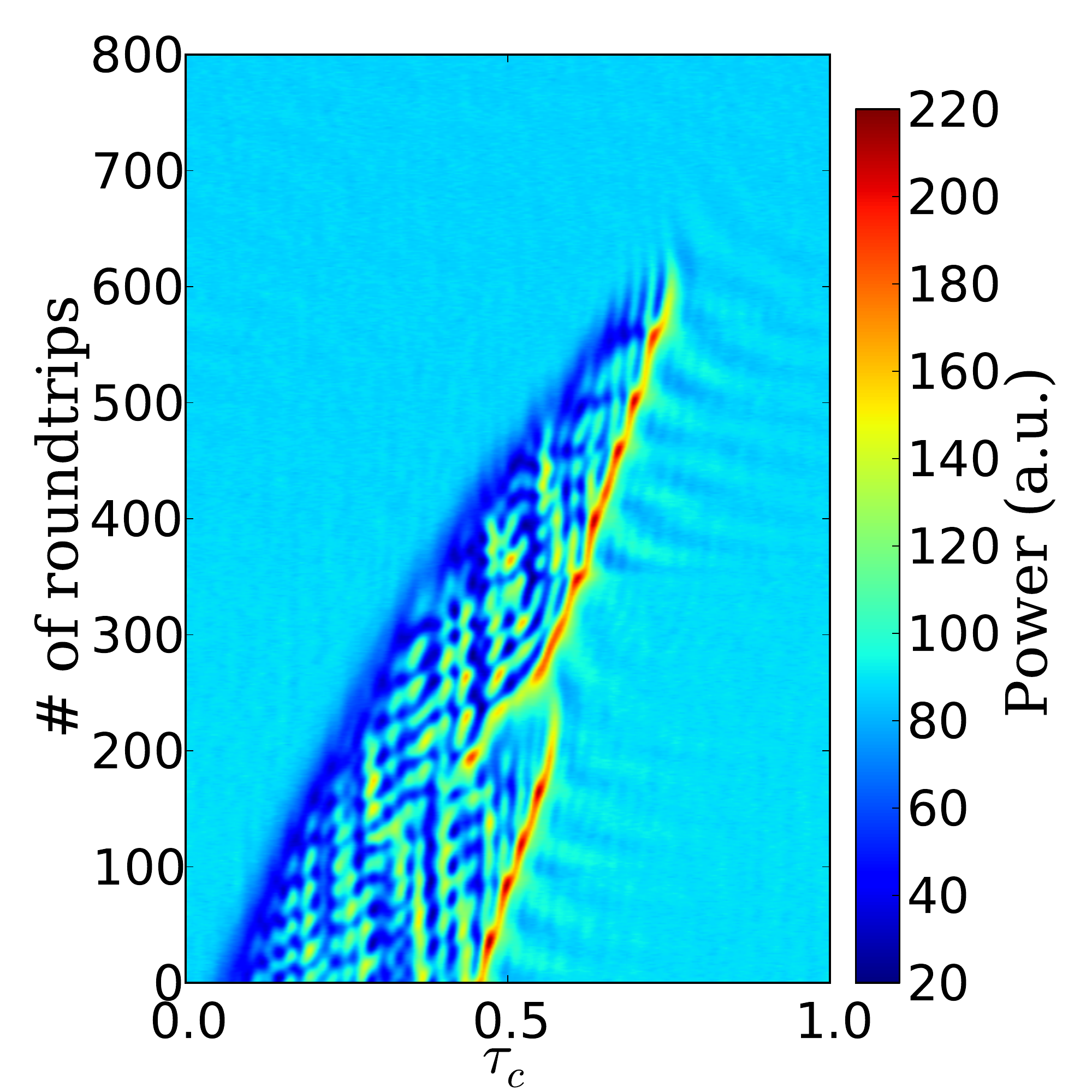}
\caption{Domain dynamics towards a phase locked regime. The chaotic domain shrinks due to different propagation speed of the left and right fronts.}
\label{coarsening}
\end{figure}

In \cite{PhysRevLett.115.043902} the chiral charge of these solitons (set by the rotation direction of the relative phase) has been shown to originate from chaotic regions, in which the electric field amplitude can vanish leading to a phase defect. Nevertheless, chaotic regions do not always carry the chiral charge which will allow soliton stability. For example in Fig. \ref{coarsening}, we show the evolution of the system towards a fully locked state without any soliton. At roundtrip 0, more than half of the system is phase locked while the remaining part is chaotic and two fronts are connecting both states. Interestingly the left and right front differ notably but inside the chaotic domain there is no obvious asymmetry. In the course of time the left and right front drift at constant but different speeds. This front propagation is strongly reminiscent of observations realized in transverse \cite{haudin2010front} and in delayed optical systems \cite{giacomelli2012coarsening}. Since no sufficient pinning force exist on the fronts \cite{verschueren2013spatiotemporal,haudin2010front,PhysRevLett.112.103901} the relative motion of the fronts leads to continuous contraction of the chaotic domain which finally disappears.

\section{Theory}

\label{sec:theory}
In order to describe field evolution in the laser with coherent injection sketched in Fig. \ref{fig:setup} we extended the model described in \cite{prati2007long,prati2010static} to include field propagation. Such a model can be easily retraced to widespread ones in multimode laser theory, incorporating the complex and peculiar optical features of semiconductor lasers and avoiding the heavy, first-principle, many-body quantum mechanical treatment. Specifically, this is done by using a phenomenological formula for the frequency and carrier density dependence of the gain and refractive index to fit the microscopic susceptibility. This leads to a simple differential equation for temporal evolution of the macroscopic polarization $P$ in the semiconductor active medium that can be coupled with the equations for the electric field $E$ and the carrier density $N$ to provide a complete and self-consistent description of the system dynamics.
\subsection{Effective semiconductor Maxwell-Bloch equations}
\label{subsec:mbe}
We suppose that both beam splitters BS have transmissivity $T \ne 0$, while all mirrors M are perfect reflectors ($T=0$).
The electric field $\tilde{E_{I}}(0,t)$ injected into the cavity at $z=0$ (the position of the right BS) and the one propagating in the active medium $\tilde{E}(z,t)$ can be written as
\begin{eqnarray}
\tilde{E}(z,t)&=&\frac{{\cal E}(z,t)}{2}\exp{\left[i(k_{0}z-\omega_{0}t)\right]}+\mathrm{c.c.}\,,\\
\tilde{E_{I}}(0,t)&=&\frac{{\cal E}_{I}}{2}\exp{(-i\omega_{0}t)}+\mathrm{c.c.}\,,
\end{eqnarray}
where ${\cal E}_{I} \in \mathbb{R}$, $k_{0}=\omega_{0}/v=\omega_{0}n/c$ and $n=\sqrt{\epsilon_{b}}$ is the background refractive index.
Analogously, the medium macroscopic polarization can be written as
\begin{equation}
\tilde{P}(z,t)=\frac{{\cal P}(z,t)}{2}\exp{\left[i(k_{0}z-\omega_{0}t)\right]}+c.c.\,.
\end{equation}
In the slowly varying envelope and rotating wave approximation, the radiation-matter interaction is described by the following nonlinear partial differential equations \cite{prati2007long}
\begin{eqnarray}
\frac{\partial{\cal E}}{\partial z}+\frac{1}{v}\frac{\partial{\cal E}}{\partial t} &=& \tilde{g}{\cal P}\,, \label{E1}\\
\tau_d\frac{\partial{\cal P}}{\partial t}&=& \left[\Gamma(N)(1-i \alpha)+2 i \delta(N)\right]\,, \label{P1} \\
&\times& \left [-if_{0}\epsilon_{0}\epsilon_{b}(1-i\alpha)\left(N/N_0-1\right){\cal E} -{\cal P}  \right]  \nonumber \\
\tau_e\frac{\partial N}{\partial t}&=&\frac{I\tau_{e}}{eV}-N-\frac{i \tau_{e}}{4 \hbar}({\cal E}^{*}{\cal P}-{\cal E}{\cal P}^{*})\,, \label{N1}
\end{eqnarray}
where $\tau_{e}$ is the carrier density nonradiative decay time, $I$ and $V$ are the pump current and the sample volume respectively,
$\tilde{g}=i \omega_{0}\Gamma_{c}/(2 \epsilon_{0}n n_{g}c)$, $n_{g}$ is the group index, $\Gamma_{c}$ is the field confinement factor, $\alpha$ is the linewidth enhancement factor, $N_{0}$ is the transparency carrier density. Finally $\Gamma(N)$ and $\delta(N)$ represent the gain width and the detuning between the gain peak and the reference frequency (as long as $|\delta(N)|\ll\Gamma(N)$) and in Eq. (\ref{P1}) the term $\zeta=f_0\epsilon_{0}\epsilon_{b}$ is the differential gain, where $f_0$ measures the maximum gain \cite{prati2007long}. The functions $\Gamma(N)$ and $\delta(N)$ can be phenomenologically derived by a linear fit of the gain curves calculated with a microscopic model for different values of $N$ \cite{prati2007long}.
In terms of the new variables $E=\eta {\cal E}$, $P'=i\eta {\cal P}$, $D=\zeta(N/N_{0}-1)$, with $\eta^{2}=\zeta\tau_{e}/(2 \hbar N_{0})$, Eqs. (\ref{E1})-(\ref{N1}) take the simplified form
\begin{eqnarray}
\frac{\partial E}{\partial z}+\frac{1}{v}\frac{\partial E}{\partial t} &=& g P\,, \label{E2}\\
\tau_d\frac{\partial P}{\partial t}&=& \left[\Gamma(D)(1-i \alpha)+2 i \delta(D)\right] \nonumber\\
&\times& \left[(1-i\alpha)E D-P \right]\,,\label{P2}   \\
\tau_e\frac{\partial D}{\partial t}&=&\mu-D-\frac{1}{2}(E^{*}P+E P^{*})\,,\label{D2}
\end{eqnarray}
where we set $g=-i\tilde{g} \, \in \mathbb{R}$, $E_I=\eta {\cal E_I}$, and $\mu=\zeta\left(I/I_0-1\right)$, being $I_0=eVN_0/\tau_e$ the transparency current. For the functions $\Gamma(D)$ and $\delta(D)$ we use the expressions reported in \cite{prati2010static} for an equivalent emitter $\Gamma(D)= 0.276+1.016 \,D$, $\delta(D)=-0.169+0.216 \,D$. To make the boundary conditions periodic and isochronous we follow the guidelines in \cite{lugiato2015nonlinear}.
The boundary condition for the field envelope at $z=0=L$ are
\begin{equation}
E(0,t)=\sqrt{T}E_{I}+RE(l,t-\Delta t)e^{-i\delta_{0}}\,. \label{BC0}
\end{equation}
where $R=1-T$, $\Delta t=(L-l)/c$, $\Lambda=L-l+nl$ and $\delta_0=(\omega_c-\omega_0)\Lambda/c$, $\omega_c$ is the cavity frequency closest to $\omega_0$ and $l$ is the length of the active medium.
By introducing the transformation $\eta=z/l$, $t'=t+\frac{z}{l}\Delta t$ the boundary condition (\ref{BC0}) assumes the isochronous form
\begin{equation}\label{BC1}
E(0,t')=\sqrt{T}E_{I}+RE(1,t')e^{-i\delta_{0}} \,.
\end{equation}
and Eq. (\ref{E2}) becomes
\begin{equation}
\frac{\partial E}{\partial \eta}+\frac{\Lambda}{c}\frac{\partial E}{\partial t'} = g P\,, \label{E21}
\end{equation}
while Eqs. (\ref{P2}) and (\ref{D2}) read the same apart from the replacement of $\partial/\partial t$ by $\partial/\partial t'$.
Finally, by introducing the new field envelopes
\begin{eqnarray}
E'(\eta,t')&=&E(z,t')\mathrm{e}^{[(\ln R-i \delta_{0})\eta]}+\sqrt{T}E_{I}\eta\,, \label{au0}\\
P'(\eta,t')&=&P(z,t')\mathrm{e}^{[(\ln R-i \delta_{0})\eta]}\,,  \label{au1}
\end{eqnarray}
we obtain
\begin{eqnarray}
\frac{\partial E'}{\partial\eta}+\frac{\Lambda}{c}\frac{\partial E'}{\partial t'} &=& \left(\ln R -i \delta_{0}\right)
\left(E' -\sqrt{T}E_{I}\eta \right) \label{E3}\\
&+&\sqrt{T}E_{I}+gl P'\,, \nonumber \\
\tau_d\frac{\partial P'}{\partial t'}&=& \left[\Gamma(D)(1-i \alpha)+2 i \delta(D)\right]\label{P3} \\
&\times& \left[(1-i\alpha)D\left(E'-\sqrt{T}E_{I}\eta\right)-P' \right]\,, \nonumber  \\
\tau_e\frac{\partial D}{\partial t'}&=&\mu-D-\frac{1}{2} \mathrm{e}^{-(2\ln R) \eta}\label{D3} \\
&\times& \left[E'^*P'+E'P'^* - \sqrt{T}E_{I}\eta\left(P'+P'^*\right)\right]\,,\nonumber
\end{eqnarray}
with the periodic and isochronous boundary condition
\begin{equation}\label{BCp}
E'(0,t')=E'(1,t')\,.
\end{equation}
At this point we apply the low transmission approximation \cite{lugiato2015nonlinear} defined as $T\ll1$, $gl\ll1$, $|\delta_{0}|\ll1$ with the pump parameter $\mathcal{A}=gl/T$ and the cavity detuning $\theta=\delta_{0}/T$ both of order unity.

In this limit the auxiliary variables $E'$ and $P'$ defined by Eqs. (\ref{au0})-(\ref{au1}) coincide with $E$ and $P$, respectively.
By introducing the dimensionless time $\tau=t'/\tau_d$ and rates $\sigma=\tau_d cT/\Lambda$, $b=\tau_d/\tau_e$, the amplitude $y=E_I/\sqrt{T}$
of the injected field, and making the substitutions $\cal{A}P\rightarrow P$, $\cal{A}D\rightarrow D$, $\cal{A}\mu\rightarrow \mu$ we can write the dynamical equations (\ref{E3})-(\ref{D3}) in the final form
\begin{eqnarray}
\frac{c\tau_{d}}{\Lambda}\frac{\partial E}{\partial \eta}+\frac{\partial E}{\partial\tau} &=&\sigma \left[y-(1+i \theta)E+P\right]\,, \label{E4}\\
\frac{\partial P}{\partial\tau}&=& \left[\Gamma(D)(1-i \alpha)+2 i \delta(D)\right] \nonumber \\
&\times& \left[(1-i\alpha)E D-P \right]\,,   \label{P4}\\
\frac{\partial D}{\partial\tau}&=&b\left [\mu-D-\frac{1}{2}\left(E^{*}P+ E P^{*} \right)\right]\,,\label{D4}
\end{eqnarray}
with $E(0,\tau)=E(1,\tau)$.  Eqs. (\ref{E4})-(\ref{D4}), apart from the propagation term, coincide with Eqs. (1)-(3) in \cite{prati2010static} if in the latter the diffraction term is neglected. The dependence of the model parameters on the medium length within the cavity $l$ is purely parametric, provided that $\sigma / b$ remains constant, since it appears only via $\Lambda$, $\eta$ and $\sigma$. This indicates that this geometrical term is relevant to the quantitative aspects of the results but not to their substance.

We observe that Maxwell-Bloch equations for multilongitudinal mode emission, analogous to Eqs. (\ref{E4})--(\ref{D4}), were introduced in \cite{PhysRevA.32.1588} for a two-level unidirectional ring laser with injected field.  In that system a multimode instability, due to the competition among the injected field frequency and the free running laser frequency was analyzed in detail.
\subsection{Reduction to rate-equations}
\label{subsec:rate}
The present model can be reduced to the widespread rate-equation model (see e.g. \cite{agrawal1993infrared}) by assuming a flat gain (infinite gain linewidth) which is formally stated by adiabatically eliminating the macroscopic polarization $P$, i.e. by setting $\frac{\partial P}{\partial \tau}=0$ in Eq. (\ref{P4})
\begin{eqnarray}
\frac{c\tau_{d}}{\Lambda}\frac{\partial E}{\partial\eta}+\frac{\partial E}{\partial\tau}&=&\sigma\left[y-(1+i\theta)E+(1-i\alpha)ED\right]\,,\label{E4r}\\
\frac{\partial D}{\partial \tau}&=&b\left [\mu-D(1+|E|^{2})\right]\,.\label{D4r}
\end{eqnarray}
 The rate-equation model has been numerically tested versus the complete one (Eqs. (\ref{E4})-(\ref{D4})) for several relevant cases and it has proved capable of describing coherent sceneries as for dynamics and stability of locked/unlocked states and phase solitons. This evidence, whose deeper analysis will be the object of further work, lead us to use the reduced approach in the systematic study of the phase solitons properties. 
\subsection{Injection locked solution and its stability}
\label{subsec:stability}
The dynamical equations (\ref{E4})-(\ref{D4}) (as well as (\ref{E4r})-(\ref{D4r})) admit the longitudinally uniform stationary solution $E=E_{s}=\sqrt{x}\mathrm{e}^{i\phi}$, $P=P_{s}$, $D=D_{s}$ (injection locked solution) with $D_s=\mu/(1+x)$, $P_s=(1-i\alpha)D_sE_s$, and
\begin{eqnarray}
y^2&=&x\left[ \left(1-D_s\right)^{2}+\left(\theta +\alpha D_s\right)^2\right]\,, \label{st3} \\
\phi& =& \arctan\left(\frac{\theta+\alpha D_s}{D_s-1}\right) \label{st4}\,.
\end{eqnarray}
Clearly $\mu=1$ is the value of the pump parameter at the free running laser threshold.

\begin{figure}[t]
\center
\includegraphics[width=0.9\columnwidth]{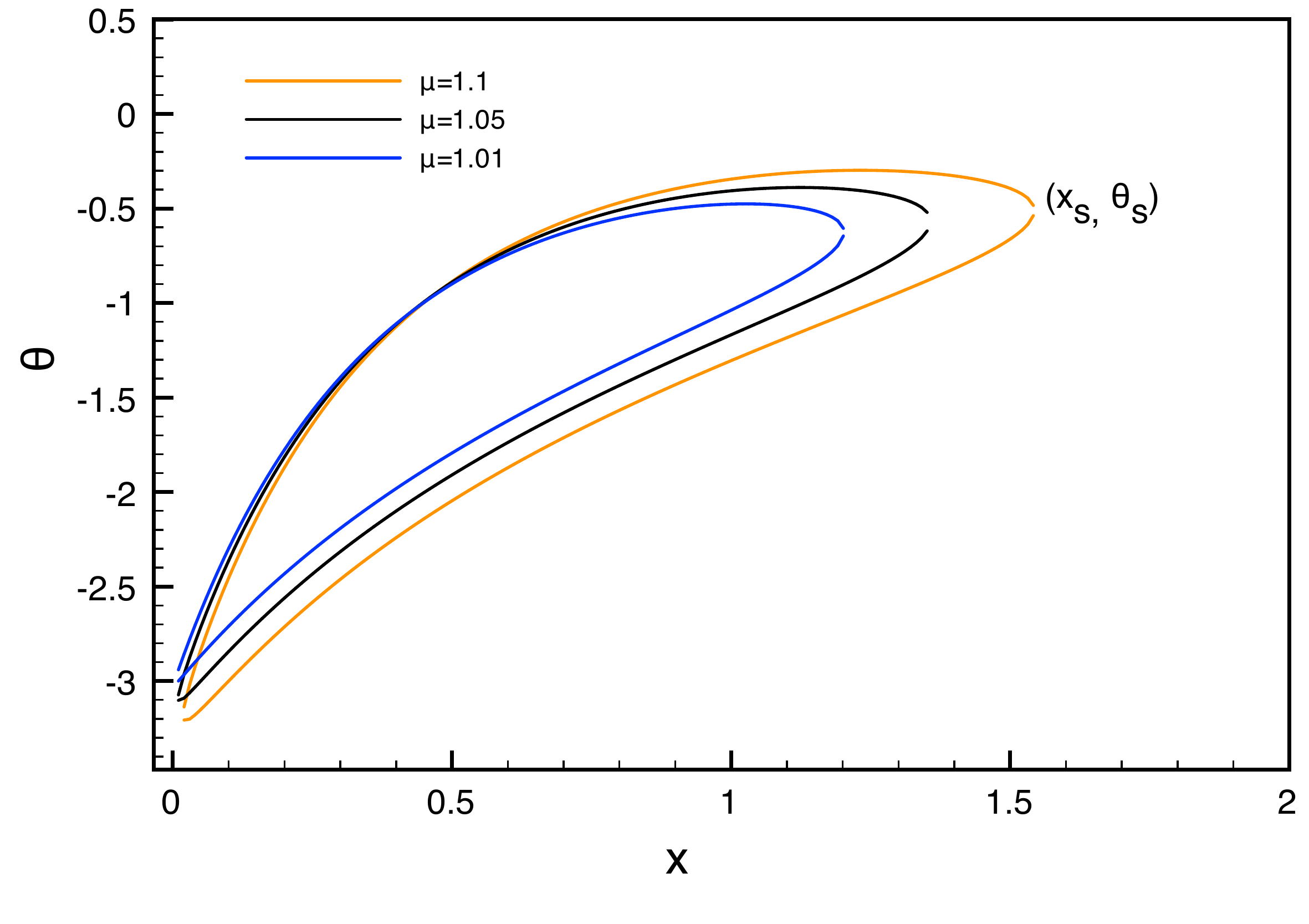}
\caption{Extension of the bistability in the $(x,\theta)$ plane for different values of $\mu$.}
\label{figc}
\end{figure}

The shape of the stationary curves given by Eqs. (\ref{st3})-(\ref{st4}) depends on the parameters $\mu$, $\alpha$, $\theta$ and it can exhibit bistability for suitable values of those parameters.
Fig. \ref{figc} shows the bistability domains in the $(x,\theta)$ plane for different values of $\mu$ and fixed $\alpha$.
The equation for the boundaries of the bistability is \cite{prati2010static}
\begin{equation}
\theta_{\pm}=-\frac{\mu\alpha\pm\sqrt{\mu^{2}x^{2} (1+\alpha^{2})-[(1+x)^{2}-\mu]^{2}}}{(1+x)^{2}}\,. \label{bist_1eq}
\end{equation}
Using Eq. (\ref{bist_1eq}) we can derive the value of $\theta_{S}$ and $x_{S}$ corresponding to the rightmost point of the bistability domain (vertical tangent) \cite{prati2010static}. In particular, one finds $\theta_{S}=-\mu\alpha/(1+x_{S})^{2}$, which implies that
in order to have bistability the detuning $\theta$ must be negative, i.e. the injection frequency must be red detuned with respect to the cavity resonance.
\begin{figure}[t]
\center
\includegraphics[width=0.98\columnwidth]{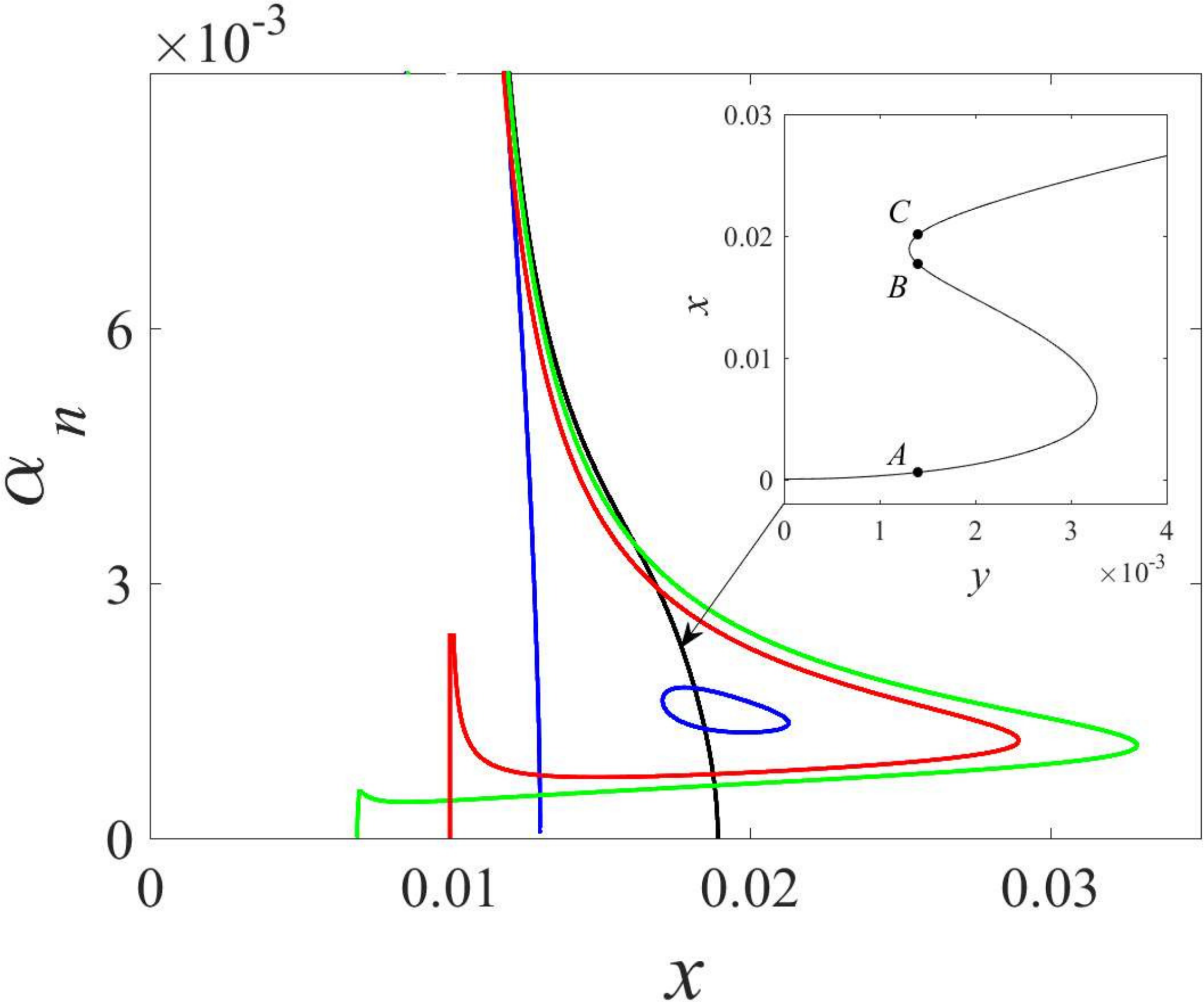}
\caption{Instability domains boundaries in the ($\alpha_{n}$, $x$) plane for $\theta=-3.01$ (green curve), $\theta=3$ (red curve), $\theta=-2.99$ (blue curve), $\theta=-2.97$ (black curve). The inset represents the S-shaped curve of the injection locked solutions corresponding to $\theta=-2.97$. The other parameters are $\alpha=3$, $\mu=1.01$, $\sigma=3\times10^{-6}$, $b=5\times10^{-4}$.}
\label{fig5}
\end{figure}

We study the stability of the stationary solution $E_{s}$, $P_{s}$, $D_{s}$ as in \cite{lugiato2015nonlinear} assuming spatio-temporal perturbations of the form $\delta X(\eta)\exp(\lambda\tau)$ where $X=E,\,E^{*},\,P,\,P^{*},\,D$, and
\begin{equation}
\delta E(\eta)=\sum_{n=-\infty}^{+\infty}\delta f_{n}e^{2\pi i n \eta}, \quad n=0,\pm 1, \pm 2 \ldots \,.\label{modal1}
\end{equation}
The linearized equations for the perturbations are
\begin{eqnarray}
\lambda\delta f_{n}&=& \left[i\alpha_{n}-\sigma(1+i\theta)\right]\delta f_{n}\nonumber\\
&&+\sigma\int_{0}^{1}d\eta\,\mathrm{e}^{-2\pi i n \eta} \delta P\,,   \label{E3modal1}\\
\lambda \delta P&=& Q \left[(1-i\alpha)(E_{s} \delta D+D_{s}\delta E) - \delta P \right]\,,   \label{P3a}\\
\lambda \delta D &=&-\frac{b}{2}\left(E_{s}\delta P+ \delta E^{*} P_{s}+E_{s}\delta P_{s}^{*}+\delta EP_{s}^{*}\right)\nonumber\\
&&-b\delta D\,,\label{P3b}
\end{eqnarray}
where $\alpha_{n}=2\pi nc \tau_{d}/\Lambda$ are the cavity resonances scaled to the gain width and we set $Q=\Gamma(D_{s})(1-i \alpha)+2 i \delta(D_{s})$.
Combining equation (\ref{P3a}) for $\delta P$ and the analogous equation for $\delta P^{*}$ (not reported here) with equation (\ref{P3b}) for $\delta D$ we get, after lengthy but simple algebra, the following expressions for $\delta P$ and $\delta P^{*}$ as a function of the electric field perturbations $\delta E$ and $\delta E^{*}$
\begin{eqnarray}
\delta P   &=& T_1 \delta E + T_2 E_s^2\delta E^*\,,\label{deltap}\\
\delta P^* &=& T_2' \left(E_s^*\right)^2\delta E + T_1' \delta E^*\,,\label{deltapstar}
\end{eqnarray}
with
\begin{eqnarray}
T_1&=&{\cal D}^{-1}\left(1-i\alpha\right)QD_s\\
&&\times\left[\left(\lambda+Q^*\right)\left(\lambda+b\right)-b\left(1+i\alpha\right)x\lambda/2\right]\,,\nonumber\\
T_2&=&-{\cal D}^{-1}\left(1-i\alpha\right)QD_sb\left[Q^*+\left(1-i\alpha\right)\lambda/2\right]\,,\\
{\cal D}&=&\lambda\left(\lambda+Q\right)\left(\lambda+Q^*\right)+b\lambda^2+2b\mathrm{Re}(Q)\lambda\nonumber\\
&&+bx\left[\mathrm{Re}(Q)+\alpha\mathrm{Im}(Q)\right]\lambda+b|Q|^2\left(1+x\right)\,.
\end{eqnarray}
The functions $T_{1,2}'$ are obtained from $T_{1,2}$ through complex conjugation but leaving $\lambda$ unaltered.
Inserting the expressions (\ref{deltap}) and (\ref{deltapstar}) in Eq. (\ref{E3modal1}) for $\delta f_{n}$ and in the corresponding equation for
$\delta f^{*}_{-n}$ we obtain the characteristic equation
\begin{eqnarray}\label{car}
&&\left[\lambda+i\alpha_n+\sigma\left(1+i\theta-T_1\right)\right]\\
&&\times\left[\lambda+i\alpha_n+\sigma\left(1-i\theta-T_1'\right)\right]-\sigma^2x^2T_2T_2'=0\,.\nonumber
\end{eqnarray}
In accordance with the experimental values, we can here safely assume $b,\,\alpha_n={\cal O}\left(\epsilon\right)$ and $\sigma={\cal O}\left(\epsilon^2\right)$ and solve Eq. (\ref{car}) perturbatively in $\epsilon$. The relevant eigenvalue can be written as $\lambda=\lambda_1\epsilon+\lambda_2\epsilon^2$, with $\lambda_1$ imaginary. Hence the instability condition is $\mathrm{Re}(\lambda_2)>0$ from which we derived an equation for the boundaries of the instabilities domain which is biquadratic in $\alpha_n$
\begin{equation}\label{boundaries}
c_4b^{-4}\alpha_n^4+c_2b^{-2}\alpha_n^2+c_0=0\,,
\end{equation}
where the coefficients $c_{i}$ are given by:
\begin{eqnarray}
c_0 &=& -\left(1+x\right)\left(1+x-D_s\right)\left[D_s^2\left(x-1\right)\left(1+\alpha^2\right)\right.\nonumber\\
&&\left.+2D_s\left(1-\alpha\theta\right)-\left(1+x\right)\left(1+\theta^2\right)\right]\,,\\
c_2 &=& -D_s^4\left(1+\alpha^2\right)\left[x^2\left(1+\alpha^2\right)-2\right]\nonumber\\
&& -2D_s^3\left[\left(2+x\right)\left(2+\alpha^2-\alpha\theta\right)\right.\nonumber\\
&& \qquad\quad\left.+x^2\left(1+\alpha^2\right)\left(\alpha\theta-1\right)\right]\nonumber\\
&& +D_s^2\left[x^2+2\left(6+\alpha^2\right)\left(1+x\right)-8\alpha\theta\left(1+x\right)\right.\nonumber\\
&& \qquad\quad\left.+2\theta^2\left(1+x\right)-\alpha^2\theta^2x^2\right]\nonumber\\
&& +2D_s\left(1+x\right)\left(2+x\right)\left(\alpha\theta-\theta^2-2\right)\nonumber\\
&& +2\left(1+\theta^2\right)\left(1+x\right)^2\,,\\
c_4 &=& \left(D_s-1\right)^2\left[\left(D_s-1\right)^2+\left(\alpha D_s+\theta\right)^2\right]\,.
\end{eqnarray}
The coefficient $c_0$ gives the boundaries of the instability domain in the single-mode limit $n=0$, and it reads
\begin{equation}
c_0=\left[(x+1)^2-\mu\right]^2\frac{dy^2}{dx}\,.
\end{equation}
If the stationary curve is S-shaped the equation $c_0=0$ is satisfied at the turning points of the curve which means that, as usual, the negative slope part of the curve is unstable. Another solution of the equation, which exists even if the stationary curve is single valued, is $x_{IL}=\sqrt{\mu}-1$, which corresponds to the injection locking threshold, i.e. the stationary solution is unstable when the injected amplitude is so small that $x<x_{IL}$.
This result agrees with that of a two-level laser, where the injection locking threshold is given by $x_{IL}=\mu-\mu_{thr}$ for a class-B laser and by
$x_{IL}=\sqrt{\mu_{thr}}\left(\sqrt{\mu}-\sqrt{\mu_{thr}}\right)$ for a class-A laser \cite{lugiato2015nonlinear}. In our case we are in the class-A limit because $\sigma\ll b$ and we have $\mu_{thr}=1$.
On the other hand, $c_4$ is always nonnegative and it is null for $D_s=1$, i.e. $x=\mu-1$. This explains the existence of a vertical asymptote in the instability domains shown in Fig. \ref{fig5} for typical cases. 

Referring to Fig. \ref{fig5}, we observe that the necessary condition for the existence of a phase soliton, that is the stability of the whole upper branch of the stationary curve, is achieved when $\theta+\alpha$ is positive and sufficiently large (black line). On the contrary if $\theta+\alpha$ is positive but very small, null, or negative a part of the upper branch in unstable against a band of sidemodes. In this case the instability domain starts from the saddle node when$\theta+\alpha$ is null or negative (red and green curves) or from a different point on the upper branch when $\theta+\alpha$ is positive. 
We will focus in the rest of the paper on the latter case $\theta=-2.97$ where the whole upper branch is stable. The corresponding S-shaped curve is shown in the figure inset together with the three fixed points for $y=0.0014$, denoted as A, B, and C.

\subsection{Numerical simulations}
\label{subsec:simulations}

We looked for phase solitons starting from point C in Fig. \ref{fig5} where the system is excitable in the absence of propagation
($\partial /\partial z=0$).%

A stable soliton is obtained by superimposing to the stable locked state $C$ a positive phase kink of $2\pi$ along $z$ for the field, having the form $\Phi_{+}(z)=4\tan^{-1}\left[\exp(-\beta z)\right]$ with $\beta$ large to have a steep kink. We checked that a stable soliton can be created in that way with any $\beta$ larger than about 5. This is to be expected, since phase solitons are robust attractors of the system. The only observed difference is in the build-up time, which is the larger the less steep is the phase jump (i.e. the smaller is $\beta$).

\begin{figure}[htb]
\includegraphics[width=0.75\columnwidth]{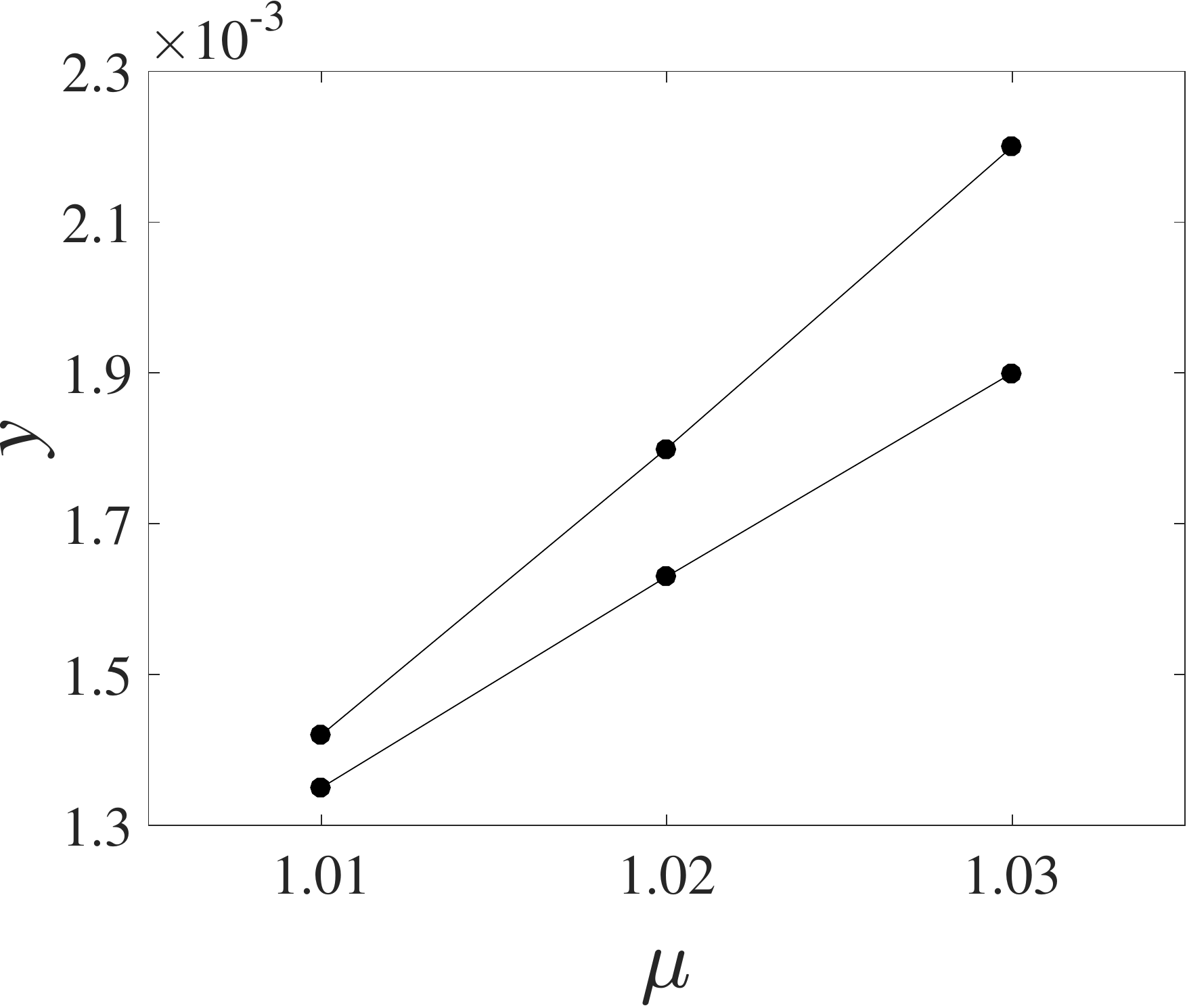}
\caption{Phase solitons existence range in terms of injected field amplitude $y$ versus the pump $\mu$. We set $\theta=-2.97$ while the other parameters are as in Fig.\ref{fig5}. }
\label{fig6}
\end{figure}

Phase solitons are stable over a finite interval of locked states $C$ which begins approximately at the left turning point of the bistable curve and whose extension increases with the pump $\mu$ as shown in Fig. \ref{fig6}.
\begin{figure}[htb]
\center
\includegraphics[width=0.26\textwidth]{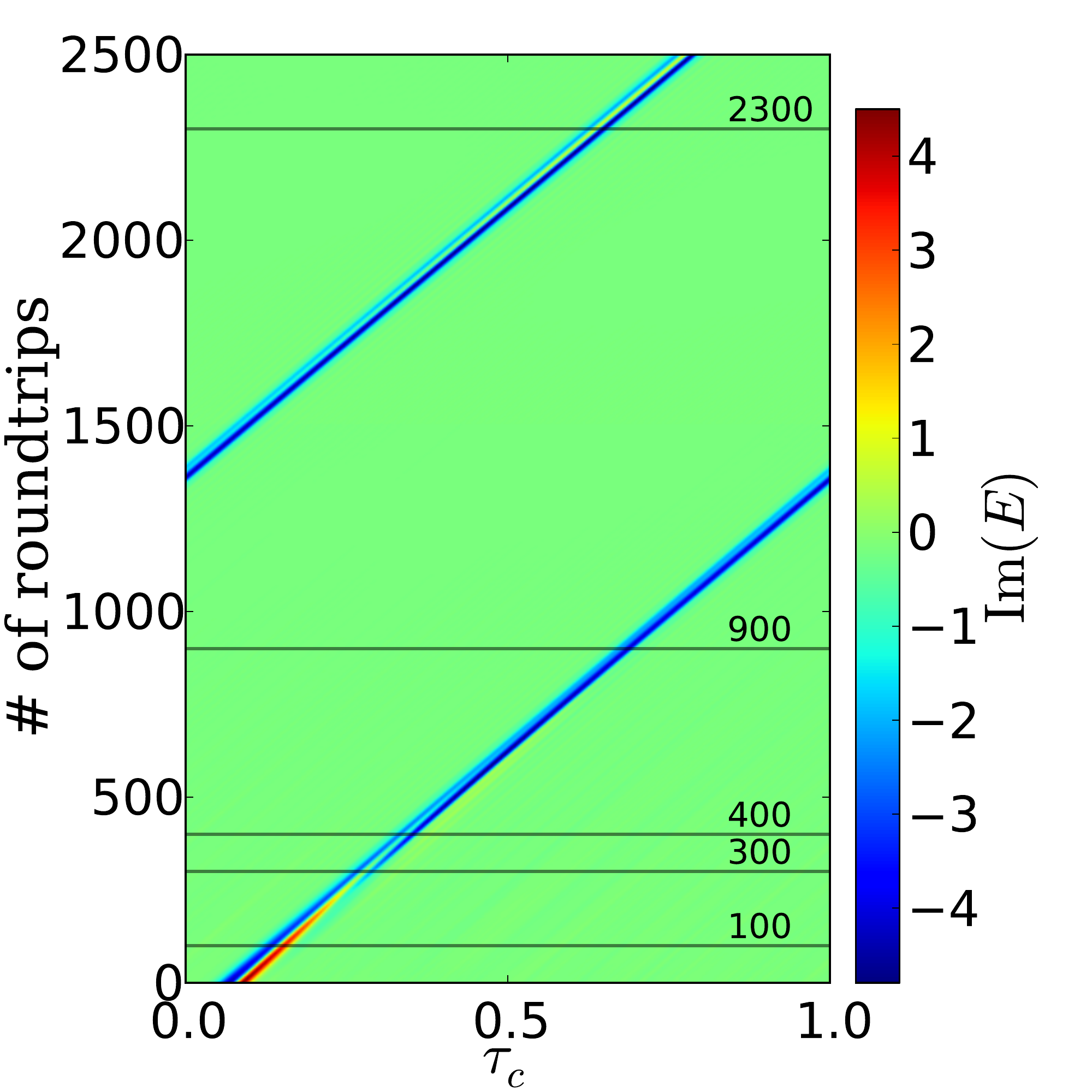}\hspace{-0.15cm}
\includegraphics[width=0.21\textwidth]{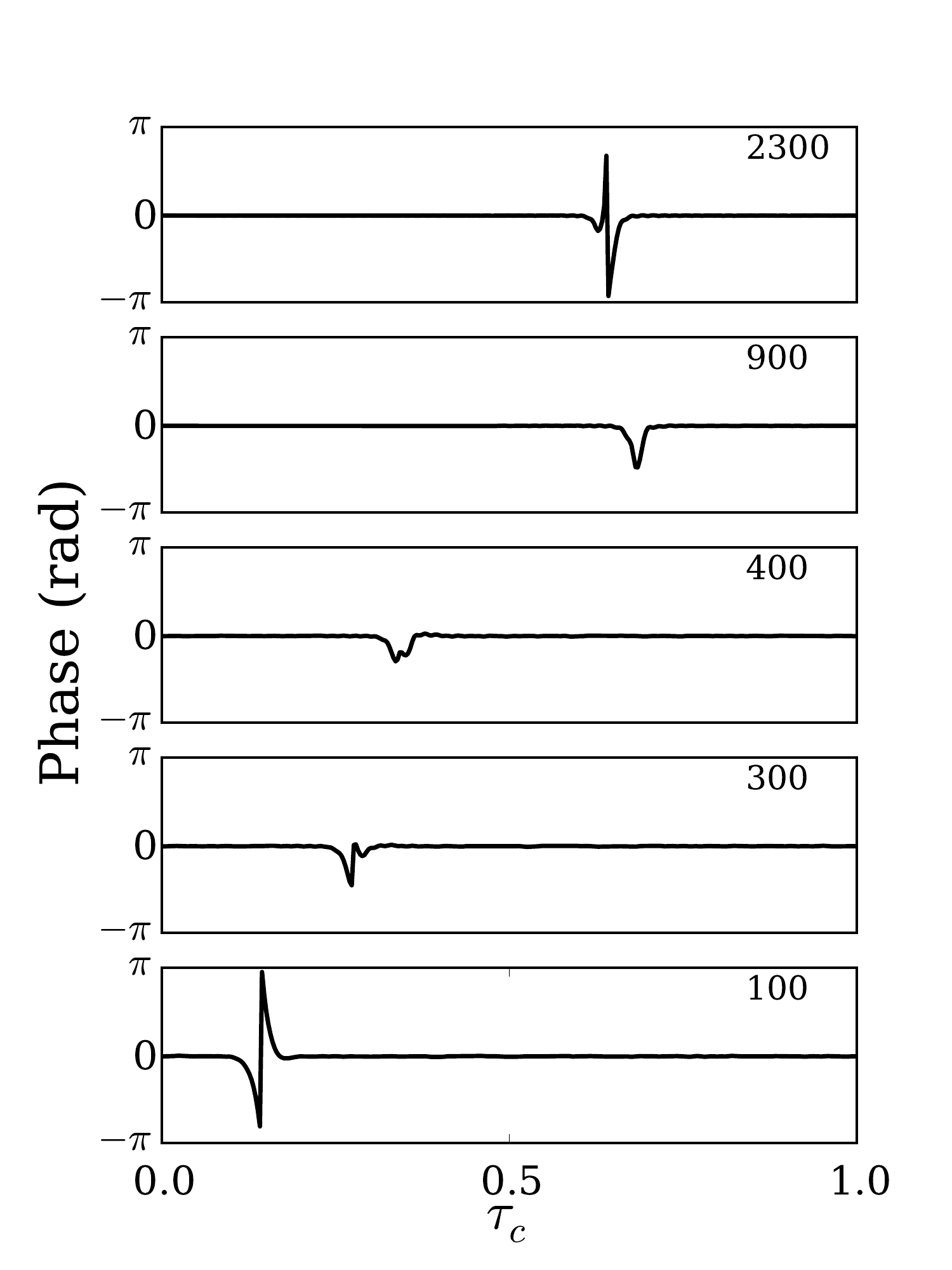}
\caption{Negatively charged solitons are unstable and acquire a positive chiral charge by passing trough a phase defect. We set $y=0.0014$ and $\theta=-2.97$. The other parameters are as in Fig.\ref{fig5}. }
\label{fig7}
\end{figure}
\begin{figure}[t]
\center
\includegraphics[width=0.45\textwidth]{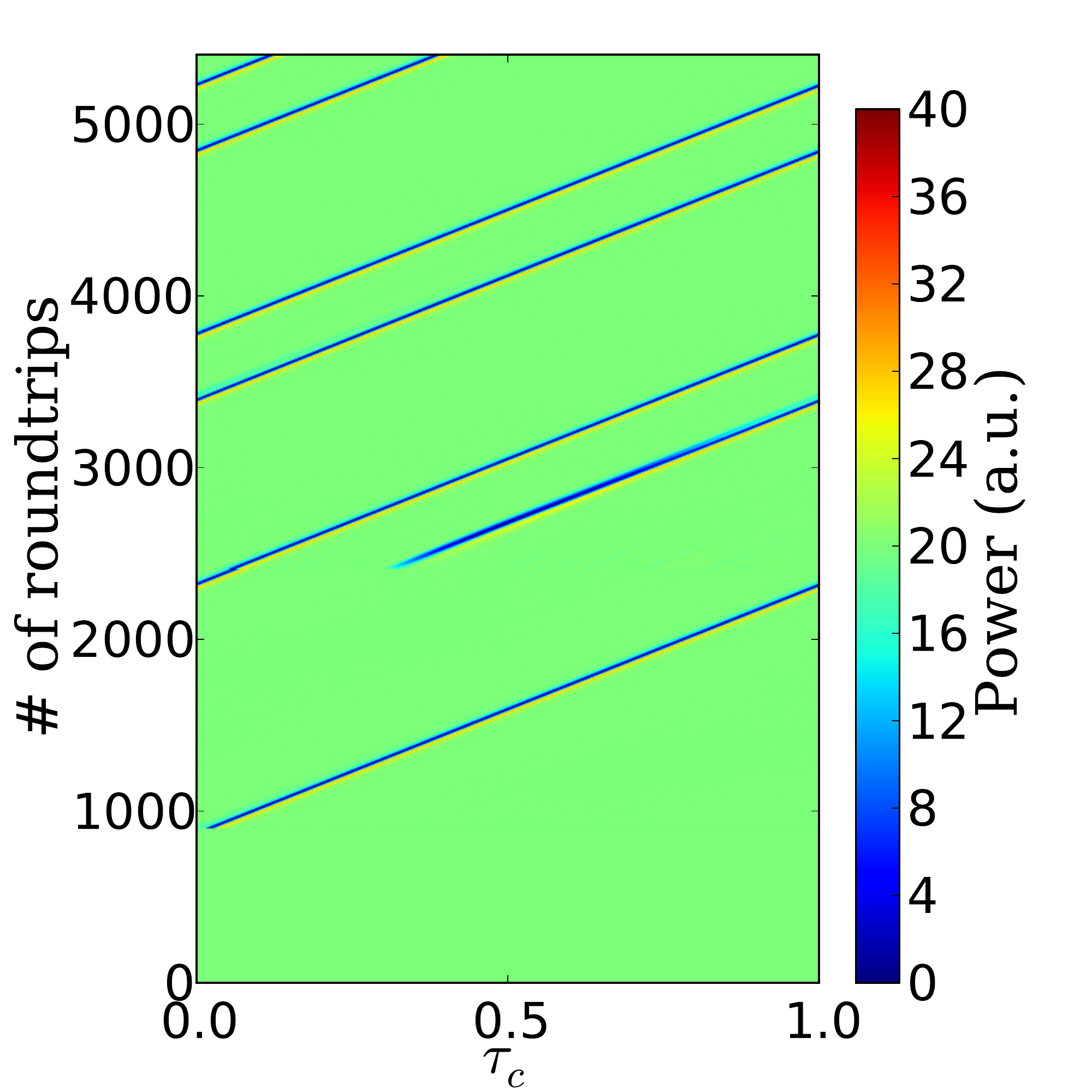}
\caption{Multistability among the homogeneous locked solution corresponding to the fixed point $A$ in Fig.\ref{fig5}, the 1 and 2 solitons configurations. The two phase solitons travel endlessly along the cavity. Parameters as in Fig.\ref{fig7}.}
\label{fig8}
\end{figure}

In order to speed up the simulations, we obtained the latter results by numerical integration of the reduced model equations Eqs. (\ref{E4r})-(\ref{D4r}), after checking their consistence with the complete model. We observe that below the left stability boundary the system falls towards the unstable lower state, where its dynamics is characterized by spatio-temporal turbulence, while above the right stability boundary the system invariably approaches the uniform stable locked state.\\

Differently from what happens in driven oscillatory systems described by the a forced GL equation \cite{chate199917} that preserves the parity symmetry, and in agreement with the experiments, in our multi longitudinal mode laser the \lq\lq negative" chiral charge is unstable. In fact, as illustrated in Fig. \ref{fig7}, the addition of a negative phase kink $\Phi_{-}(z)=2\pi-4\tan^{-1}\left[\exp(-\beta z)\right]$ to the field phase excites a negative charged phase soliton that after a rather short transient, spontaneously undergoes a phase kink sign reversal. This is accomplished with by a passage of the field through the origin ($E=0\Longleftrightarrow Im(E)=Re(E)=0$) in the phase space trajectory (see Fig. \ref{solshape}.b), which is associated with the creation of a transient phase defect.

Very interestingly for applications to all optical intensity \cite{solitette,temporalcs} or phase \cite{garbin2015topological} information encoding, the multistability among the homogeneous locked state, the 1 and the 2 phase solitons solutions was demonstrated by superimposing to the stable homogeneous background first a positive kink and then a second one centered in a different position (see Fig. \ref{fig8}).

\begin{figure}[t]
\center
\includegraphics[width=0.45\textwidth]{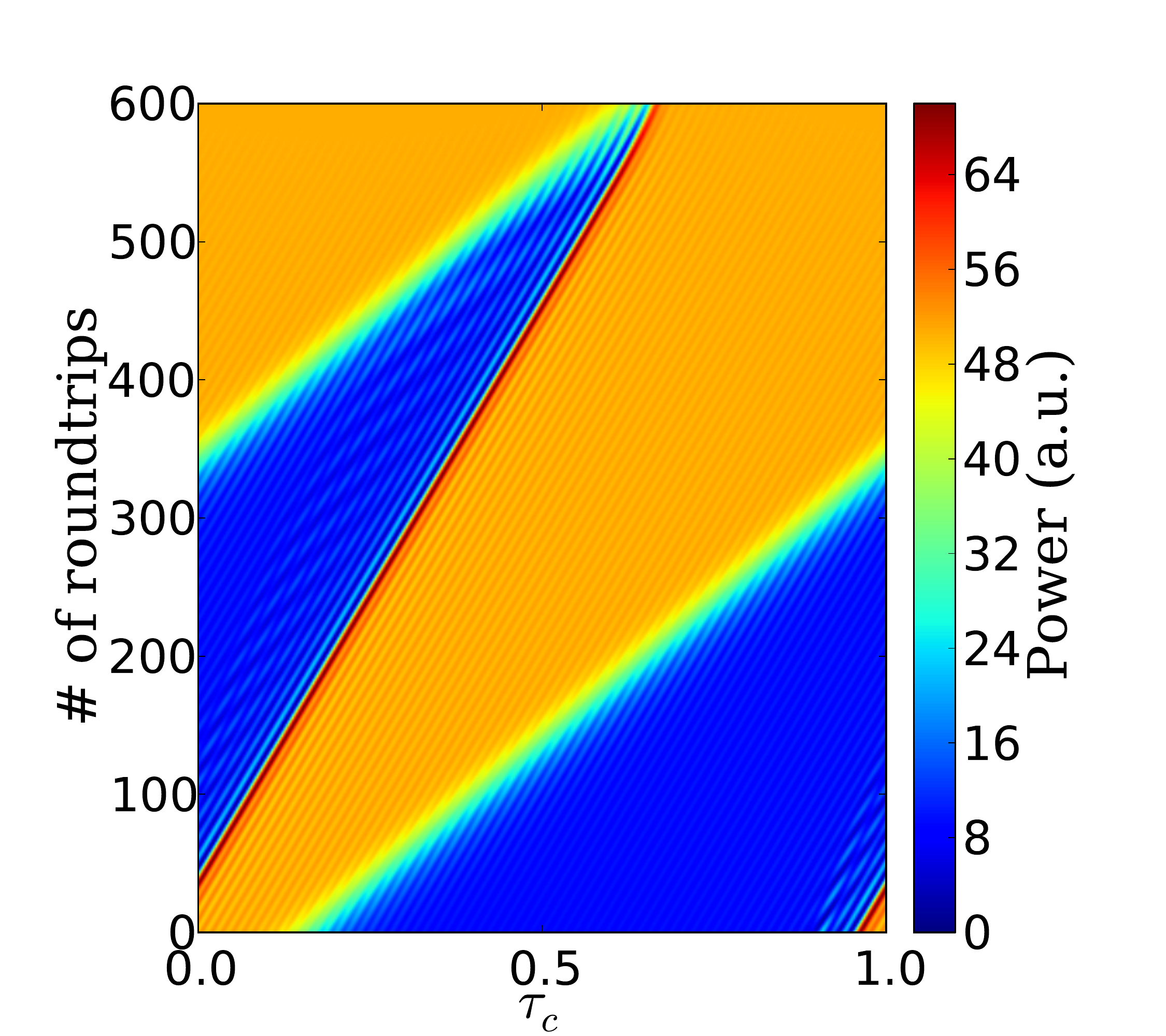}
\caption{Domain coarsening. Fronts between stable locked states and unstable states move at constant speed. The paremeters are: $\alpha=3$, $\mu=1.01$, $\sigma=3\times10^{-6}$, $b=5\times10^{-4}$, $\theta=-2.0$, $y=0.235$.}
\label{fig:fronts_numerics}
\end{figure}

Finally, one of the striking experimental observations is the asymmetry between right and left fronts shown in Fig. \ref{coarsening}. This feature, which was not observed in numerical simulations of the forced Ginzburg-Landau equation \cite{PhysRevLett.115.043902}, is on the contrary clearly visible when simulating the full model or the rate-equation model as shown in Fig. \ref{fig:fronts_numerics} and then we conclude that it is not related to the propagative nature of the system but to the non-instantaneous semiconductor medium. 
Similarly to the experimental observation, the growth of the locked state seems to result from a convective instability since this state drifts towards the right faster than it grows. However, the distinction is of course strongly related to the chosen reference frame \cite{van2003front} and in the present case of periodic boundary conditions the stable state ends up invading the whole system in any case. 

\section{Conclusions}
\label{sec:conclusions}
We have presented an experimental, theoretical and numerical study of a semiconductor laser with coherent optical injection. At variance with most previous works on the topic, we focus on a strongly multimode regime and the resulting spatio-temporal effects. Thanks to a balanced choice of the geometrical parameters of the experiment, we are able to observe long term dynamics still resolving the most detailed features. In particular we have presented observations of plane-wave and modulational instabilities. In a suitable spatio-temporal representation we have been able to describe complex time series in terms of motion of fronts connecting chaotic and phase locked domains. When these chaotic domains contain a chiral charge, the inward motion of fronts is halted and chirally charged solitons emerge. We have measured the propagation of these phase solitons over 40~km distance and therefore shown their extreme robustness. Most of these observations can be understood from the general literature in forced oscillatory media, but at least two very unexpected features could be described only with the use of a more specific modeling: phase solitons are observed for only one sign of chiral charge and left and right fronts differ strongly.
The experimental results are paired by a detailed derivation of the model sketched in \cite{PhysRevLett.115.043902} where a linear stability of the homogeneous phase locked solutions was performed. This model was further reduced to a set of rate-equations including propagation which are still able to capture all the specific features of the experiment, contrary to the oversimplified approximation of an instantaneous medium response. From this, we conclude that the asymmetry of the fronts and the instability of one of the two possible chiral signs results from the non-instantaneous medium dynamics. In order to clarify this point, experiments based on the propagation of light in instantaneous media would certainely be insightful. In particular, the analysis of rising and falling fronts in the regime of coexistence of pattern and homogeneous state outside the front pinning region in coherently driven Kerr ring resonators \cite{jang2013ultraweak,erkintalo2015bunching} would be particularly interesting. On the other hand, theoretical analysis dedicated to the stability of chirally charged walls \cite{gomila2015theory} might also shed more light on the origin of the instability of the negatively charged solitons. In addition, the destabilization of solitons (which is hard to address with our Partial Differential Equation model) might be more tractable with models based on  Delayed Differential Equations  \cite{vladimirov2005model,vladimirov2014modeling}, whose analysis could therefore prove very insightful.

Finally, our models predict that the pulse width associated to the intensity shape of the phase solitons could be significantly reduced to the ps of sub-ps scale if a shorter cavity is considered and the medium response is made correspondingly faster to ensure a formally analogue dynamical behavior in the multimode regime.

\begin{acknowledgments}
The authors would like to thank Drs. L.~Gil and M.~Argentina for many helpful discussions. LC acknowledges financial support from the MIUR project number PON02-0576. FG, GT and SB acknowledge funding from Agence Nationale de la Recherche through grant number ANR-12-JS04-0002-01.
\end{acknowledgments}

\bibliography{phaselong}

\begin{thebibliography}{35}
\expandafter\ifx\csname natexlab\endcsname\relax\def\natexlab#1{#1}\fi
\expandafter\ifx\csname bibnamefont\endcsname\relax
  \def\bibnamefont#1{#1}\fi
\expandafter\ifx\csname bibfnamefont\endcsname\relax
  \def\bibfnamefont#1{#1}\fi
\expandafter\ifx\csname citenamefont\endcsname\relax
  \def\citenamefont#1{#1}\fi
\expandafter\ifx\csname url\endcsname\relax
  \def\url#1{\texttt{#1}}\fi
\expandafter\ifx\csname urlprefix\endcsname\relax\def\urlprefix{URL }\fi
\providecommand{\bibinfo}[2]{#2}
\providecommand{\eprint}[2][]{\url{#2}}

\bibitem[{\citenamefont{Wieczorek et~al.}(2005)\citenamefont{Wieczorek,
  Krauskopf, Simpson, and Lenstra}}]{Wieczorek2005}
\bibinfo{author}{\bibfnamefont{S.}~\bibnamefont{Wieczorek}},
  \bibinfo{author}{\bibfnamefont{B.}~\bibnamefont{Krauskopf}},
  \bibinfo{author}{\bibfnamefont{T.}~\bibnamefont{Simpson}}, \bibnamefont{and}
  \bibinfo{author}{\bibfnamefont{D.}~\bibnamefont{Lenstra}},
  \bibinfo{journal}{Phys. Rep.} \textbf{\bibinfo{volume}{416}},
  \bibinfo{pages}{1} (\bibinfo{year}{2005}).

\bibitem[{\citenamefont{Lugiato et~al.}(2015)\citenamefont{Lugiato, Prati, and
  Brambilla}}]{lugiato2015nonlinear}
\bibinfo{author}{\bibfnamefont{L.}~\bibnamefont{Lugiato}},
  \bibinfo{author}{\bibfnamefont{F.}~\bibnamefont{Prati}}, \bibnamefont{and}
  \bibinfo{author}{\bibfnamefont{M.}~\bibnamefont{Brambilla}},
  \emph{\bibinfo{title}{Nonlinear Optical Systems}}
  (\bibinfo{publisher}{Cambridge University Press}, \bibinfo{year}{2015}).

\bibitem[{\citenamefont{Arnold}(1989)}]{arnold1989mathematical}
\bibinfo{author}{\bibfnamefont{V.~I.} \bibnamefont{Arnold}},
  \emph{\bibinfo{title}{Mathematical methods of classical mechanics}},
  vol.~\bibinfo{volume}{60} (\bibinfo{publisher}{Springer Science \& Business
  Media}, \bibinfo{year}{1989}).

\bibitem[{\citenamefont{Kelleher et~al.}(2009)\citenamefont{Kelleher, Goulding,
  Hegarty, Huyet, Cong, Martinez, Lema{\^{i}}tre, Ramdane, Fischer,
  Gersch{\"{u}}tz et~al.}}]{kelleher2009excitable}
\bibinfo{author}{\bibfnamefont{B.}~\bibnamefont{Kelleher}},
  \bibinfo{author}{\bibfnamefont{D.}~\bibnamefont{Goulding}},
  \bibinfo{author}{\bibfnamefont{S.~P.} \bibnamefont{Hegarty}},
  \bibinfo{author}{\bibfnamefont{G.}~\bibnamefont{Huyet}},
  \bibinfo{author}{\bibfnamefont{D.-Y.} \bibnamefont{Cong}},
  \bibinfo{author}{\bibfnamefont{A.}~\bibnamefont{Martinez}},
  \bibinfo{author}{\bibfnamefont{A.}~\bibnamefont{Lema{\^{i}}tre}},
  \bibinfo{author}{\bibfnamefont{A.}~\bibnamefont{Ramdane}},
  \bibinfo{author}{\bibfnamefont{M.}~\bibnamefont{Fischer}},
  \bibinfo{author}{\bibfnamefont{F.}~\bibnamefont{Gersch{\"{u}}tz}},
  \bibnamefont{et~al.}, \bibinfo{journal}{Optics letters}
  \textbf{\bibinfo{volume}{34}}, \bibinfo{pages}{440} (\bibinfo{year}{2009}).

\bibitem[{\citenamefont{Vaudel et~al.}(2008)\citenamefont{Vaudel, Peraud, and
  Besnard}}]{vaudel2008synchronization}
\bibinfo{author}{\bibfnamefont{O.}~\bibnamefont{Vaudel}},
  \bibinfo{author}{\bibfnamefont{N.}~\bibnamefont{Peraud}}, \bibnamefont{and}
  \bibinfo{author}{\bibfnamefont{P.}~\bibnamefont{Besnard}}, in
  \emph{\bibinfo{booktitle}{Photonics Europe}} (\bibinfo{year}{2008}), pp.
  \bibinfo{pages}{69970--69970}.

\bibitem[{\citenamefont{Kelleher et~al.}(2010)\citenamefont{Kelleher, Goulding,
  Baselga~Pascual, Hegarty, and Huyet}}]{Kelleher2010}
\bibinfo{author}{\bibfnamefont{B.}~\bibnamefont{Kelleher}},
  \bibinfo{author}{\bibfnamefont{D.}~\bibnamefont{Goulding}},
  \bibinfo{author}{\bibfnamefont{B.}~\bibnamefont{Baselga~Pascual}},
  \bibinfo{author}{\bibfnamefont{S.~P.} \bibnamefont{Hegarty}},
  \bibnamefont{and} \bibinfo{author}{\bibfnamefont{G.}~\bibnamefont{Huyet}},
  \bibinfo{journal}{The European Physical Journal D}
  \textbf{\bibinfo{volume}{58}}, \bibinfo{pages}{175} (\bibinfo{year}{2010}),
  ISSN \bibinfo{issn}{1434-6060}.

\bibitem[{\citenamefont{Turconi et~al.}(2013)\citenamefont{Turconi, Garbin,
  Feyereisen, Giudici, and Barland}}]{turconi2013control}
\bibinfo{author}{\bibfnamefont{M.}~\bibnamefont{Turconi}},
  \bibinfo{author}{\bibfnamefont{B.}~\bibnamefont{Garbin}},
  \bibinfo{author}{\bibfnamefont{M.}~\bibnamefont{Feyereisen}},
  \bibinfo{author}{\bibfnamefont{M.}~\bibnamefont{Giudici}}, \bibnamefont{and}
  \bibinfo{author}{\bibfnamefont{S.}~\bibnamefont{Barland}},
  \bibinfo{journal}{Physical Review E} \textbf{\bibinfo{volume}{88}},
  \bibinfo{pages}{022923} (\bibinfo{year}{2013}).

\bibitem[{\citenamefont{Coullet et~al.}(1998)\citenamefont{Coullet, Daboussy,
  and Tredicce}}]{coulletexcitwaves}
\bibinfo{author}{\bibfnamefont{P.}~\bibnamefont{Coullet}},
  \bibinfo{author}{\bibfnamefont{D.}~\bibnamefont{Daboussy}}, \bibnamefont{and}
  \bibinfo{author}{\bibfnamefont{J.~R.} \bibnamefont{Tredicce}},
  \bibinfo{journal}{Phys. Rev. E} \textbf{\bibinfo{volume}{58}},
  \bibinfo{pages}{5347} (\bibinfo{year}{1998}).

\bibitem[{\citenamefont{Castelli et~al.}(1994)\citenamefont{Castelli, Lugiato,
  and Pirovano}}]{Castelli1994}
\bibinfo{author}{\bibfnamefont{F.}~\bibnamefont{Castelli}},
  \bibinfo{author}{\bibfnamefont{L.}~\bibnamefont{Lugiato}}, \bibnamefont{and}
  \bibinfo{author}{\bibfnamefont{R.}~\bibnamefont{Pirovano}},
  \bibinfo{journal}{Phys. Rev. A} \textbf{\bibinfo{volume}{49}},
  \bibinfo{pages}{4031} (\bibinfo{year}{1994}).

\bibitem[{\citenamefont{Narducci et~al.}(1985)\citenamefont{Narducci, Tredicce,
  Lugiato, Abraham, and Bandy}}]{PhysRevA.32.1588}
\bibinfo{author}{\bibfnamefont{L.~M.} \bibnamefont{Narducci}},
  \bibinfo{author}{\bibfnamefont{J.~R.} \bibnamefont{Tredicce}},
  \bibinfo{author}{\bibfnamefont{L.~A.} \bibnamefont{Lugiato}},
  \bibinfo{author}{\bibfnamefont{N.~B.} \bibnamefont{Abraham}},
  \bibnamefont{and} \bibinfo{author}{\bibfnamefont{D.~K.} \bibnamefont{Bandy}},
  \bibinfo{journal}{Phys. Rev. A} \textbf{\bibinfo{volume}{32}},
  \bibinfo{pages}{1588} (\bibinfo{year}{1985}).

\bibitem[{\citenamefont{Longhi}(1998)}]{315223}
\bibinfo{author}{\bibfnamefont{S.}~\bibnamefont{Longhi}},
  \bibinfo{journal}{Quantum and Semiclassical Optics: Journal of the European
  Optical Society Part B (1995-1998)} \textbf{\bibinfo{volume}{10}},
  \bibinfo{pages}{617} (\bibinfo{year}{1998}), ISSN \bibinfo{issn}{1355-5111}.

\bibitem[{\citenamefont{Coullet}(1986)}]{PhysRevLett.56.724}
\bibinfo{author}{\bibfnamefont{P.}~\bibnamefont{Coullet}},
  \bibinfo{journal}{Phys. Rev. Lett.} \textbf{\bibinfo{volume}{56}},
  \bibinfo{pages}{724} (\bibinfo{year}{1986}).

\bibitem[{\citenamefont{Coullet and Emilsson}(1992)}]{coullet1992strong}
\bibinfo{author}{\bibfnamefont{P.}~\bibnamefont{Coullet}} \bibnamefont{and}
  \bibinfo{author}{\bibfnamefont{K.}~\bibnamefont{Emilsson}},
  \bibinfo{journal}{Physica D: Nonlinear Phenomena}
  \textbf{\bibinfo{volume}{61}}, \bibinfo{pages}{119} (\bibinfo{year}{1992}).

\bibitem[{\citenamefont{Chat{\'{e}} et~al.}(1999)\citenamefont{Chat{\'{e}},
  Pikovsky, and Rudzick}}]{chate199917}
\bibinfo{author}{\bibfnamefont{H.}~\bibnamefont{Chat{\'{e}}}},
  \bibinfo{author}{\bibfnamefont{A.}~\bibnamefont{Pikovsky}}, \bibnamefont{and}
  \bibinfo{author}{\bibfnamefont{O.}~\bibnamefont{Rudzick}},
  \bibinfo{journal}{Physica D: Nonlinear Phenomena}
  \textbf{\bibinfo{volume}{131}}, \bibinfo{pages}{17} (\bibinfo{year}{1999}).

\bibitem[{\citenamefont{Gustave et~al.}(2015)\citenamefont{Gustave, Columbo,
  Tissoni, Brambilla, Prati, Kelleher, Tykalewicz, and
  Barland}}]{PhysRevLett.115.043902}
\bibinfo{author}{\bibfnamefont{F.}~\bibnamefont{Gustave}},
  \bibinfo{author}{\bibfnamefont{L.}~\bibnamefont{Columbo}},
  \bibinfo{author}{\bibfnamefont{G.}~\bibnamefont{Tissoni}},
  \bibinfo{author}{\bibfnamefont{M.}~\bibnamefont{Brambilla}},
  \bibinfo{author}{\bibfnamefont{F.}~\bibnamefont{Prati}},
  \bibinfo{author}{\bibfnamefont{B.}~\bibnamefont{Kelleher}},
  \bibinfo{author}{\bibfnamefont{B.}~\bibnamefont{Tykalewicz}},
  \bibnamefont{and} \bibinfo{author}{\bibfnamefont{S.}~\bibnamefont{Barland}},
  \bibinfo{journal}{Phys. Rev. Lett.} \textbf{\bibinfo{volume}{115}},
  \bibinfo{pages}{043902} (\bibinfo{year}{2015}).

\bibitem[{\citenamefont{Tierno et~al.}(2012)\citenamefont{Tierno, Gustave, and
  Barland}}]{Tierno:12}
\bibinfo{author}{\bibfnamefont{A.}~\bibnamefont{Tierno}},
  \bibinfo{author}{\bibfnamefont{F.}~\bibnamefont{Gustave}}, \bibnamefont{and}
  \bibinfo{author}{\bibfnamefont{S.}~\bibnamefont{Barland}},
  \bibinfo{journal}{Opt. Lett.} \textbf{\bibinfo{volume}{37}},
  \bibinfo{pages}{2004} (\bibinfo{year}{2012}).

\bibitem[{\citenamefont{Barland et~al.}(2003)\citenamefont{Barland, Piro,
  Giudici, Tredicce, and Balle}}]{prethermal}
\bibinfo{author}{\bibfnamefont{S.}~\bibnamefont{Barland}},
  \bibinfo{author}{\bibfnamefont{O.}~\bibnamefont{Piro}},
  \bibinfo{author}{\bibfnamefont{M.}~\bibnamefont{Giudici}},
  \bibinfo{author}{\bibfnamefont{J.~R.} \bibnamefont{Tredicce}},
  \bibnamefont{and} \bibinfo{author}{\bibfnamefont{S.}~\bibnamefont{Balle}},
  \bibinfo{journal}{Phys. Rev. E} \textbf{\bibinfo{volume}{68}},
  \bibinfo{pages}{036209} (\bibinfo{year}{2003}).

\bibitem[{\citenamefont{Spinelli et~al.}(2002)\citenamefont{Spinelli, Tissoni,
  Lugiato, and Brambilla}}]{spinelli2002thermal}
\bibinfo{author}{\bibfnamefont{L.}~\bibnamefont{Spinelli}},
  \bibinfo{author}{\bibfnamefont{G.}~\bibnamefont{Tissoni}},
  \bibinfo{author}{\bibfnamefont{L.~A.} \bibnamefont{Lugiato}},
  \bibnamefont{and}
  \bibinfo{author}{\bibfnamefont{M.}~\bibnamefont{Brambilla}},
  \bibinfo{journal}{Physical Review A} \textbf{\bibinfo{volume}{66}},
  \bibinfo{pages}{023817} (\bibinfo{year}{2002}).

\bibitem[{\citenamefont{Tissoni et~al.}(2002)\citenamefont{Tissoni, Spinelli,
  Lugiato, Brambilla, Perrini, and Maggipinto}}]{tissoni2002spatio}
\bibinfo{author}{\bibfnamefont{G.}~\bibnamefont{Tissoni}},
  \bibinfo{author}{\bibfnamefont{L.}~\bibnamefont{Spinelli}},
  \bibinfo{author}{\bibfnamefont{L.}~\bibnamefont{Lugiato}},
  \bibinfo{author}{\bibfnamefont{M.}~\bibnamefont{Brambilla}},
  \bibinfo{author}{\bibfnamefont{I.}~\bibnamefont{Perrini}}, \bibnamefont{and}
  \bibinfo{author}{\bibfnamefont{T.}~\bibnamefont{Maggipinto}},
  \bibinfo{journal}{Optics express} \textbf{\bibinfo{volume}{10}},
  \bibinfo{pages}{1009} (\bibinfo{year}{2002}).

\bibitem[{\citenamefont{Haudin et~al.}(2010)\citenamefont{Haudin, El{\'{i}}as,
  Rojas, Bortolozzo, Clerc, and Residori}}]{haudin2010front}
\bibinfo{author}{\bibfnamefont{F.}~\bibnamefont{Haudin}},
  \bibinfo{author}{\bibfnamefont{R.~G.} \bibnamefont{El{\'{i}}as}},
  \bibinfo{author}{\bibfnamefont{R.~G.} \bibnamefont{Rojas}},
  \bibinfo{author}{\bibfnamefont{U.}~\bibnamefont{Bortolozzo}},
  \bibinfo{author}{\bibfnamefont{M.~G.} \bibnamefont{Clerc}}, \bibnamefont{and}
  \bibinfo{author}{\bibfnamefont{S.}~\bibnamefont{Residori}},
  \bibinfo{journal}{Physical Review E} \textbf{\bibinfo{volume}{81}},
  \bibinfo{pages}{056203} (\bibinfo{year}{2010}).

\bibitem[{\citenamefont{Giacomelli et~al.}(2012)\citenamefont{Giacomelli,
  Marino, Zaks, and Yanchuk}}]{giacomelli2012coarsening}
\bibinfo{author}{\bibfnamefont{G.}~\bibnamefont{Giacomelli}},
  \bibinfo{author}{\bibfnamefont{F.}~\bibnamefont{Marino}},
  \bibinfo{author}{\bibfnamefont{M.~A.} \bibnamefont{Zaks}}, \bibnamefont{and}
  \bibinfo{author}{\bibfnamefont{S.}~\bibnamefont{Yanchuk}},
  \bibinfo{journal}{EPL (Europhysics Letters)} \textbf{\bibinfo{volume}{99}},
  \bibinfo{pages}{58005} (\bibinfo{year}{2012}).

\bibitem[{\citenamefont{Verschueren et~al.}(2013)\citenamefont{Verschueren,
  Bortolozzo, Clerc, and Residori}}]{verschueren2013spatiotemporal}
\bibinfo{author}{\bibfnamefont{N.}~\bibnamefont{Verschueren}},
  \bibinfo{author}{\bibfnamefont{U.}~\bibnamefont{Bortolozzo}},
  \bibinfo{author}{\bibfnamefont{M.~G.} \bibnamefont{Clerc}}, \bibnamefont{and}
  \bibinfo{author}{\bibfnamefont{S.}~\bibnamefont{Residori}},
  \bibinfo{journal}{Physical review letters} \textbf{\bibinfo{volume}{110}},
  \bibinfo{pages}{104101} (\bibinfo{year}{2013}).

\bibitem[{\citenamefont{Marino et~al.}(2014)\citenamefont{Marino, Giacomelli,
  and Barland}}]{PhysRevLett.112.103901}
\bibinfo{author}{\bibfnamefont{F.}~\bibnamefont{Marino}},
  \bibinfo{author}{\bibfnamefont{G.}~\bibnamefont{Giacomelli}},
  \bibnamefont{and} \bibinfo{author}{\bibfnamefont{S.}~\bibnamefont{Barland}},
  \bibinfo{journal}{Phys. Rev. Lett.} \textbf{\bibinfo{volume}{112}},
  \bibinfo{pages}{103901} (\bibinfo{year}{2014}).

\bibitem[{\citenamefont{Prati and Columbo}(2007)}]{prati2007long}
\bibinfo{author}{\bibfnamefont{F.}~\bibnamefont{Prati}} \bibnamefont{and}
  \bibinfo{author}{\bibfnamefont{L.}~\bibnamefont{Columbo}},
  \bibinfo{journal}{Physical Review A} \textbf{\bibinfo{volume}{75}},
  \bibinfo{pages}{053811} (\bibinfo{year}{2007}).

\bibitem[{\citenamefont{Prati et~al.}(2010)\citenamefont{Prati, Tissoni,
  McIntyre, and Oppo}}]{prati2010static}
\bibinfo{author}{\bibfnamefont{F.}~\bibnamefont{Prati}},
  \bibinfo{author}{\bibfnamefont{G.}~\bibnamefont{Tissoni}},
  \bibinfo{author}{\bibfnamefont{C.}~\bibnamefont{McIntyre}}, \bibnamefont{and}
  \bibinfo{author}{\bibfnamefont{G.~L.} \bibnamefont{Oppo}},
  \bibinfo{journal}{The European Physical Journal D-Atomic, Molecular, Optical
  and Plasma Physics} \textbf{\bibinfo{volume}{59}}, \bibinfo{pages}{139}
  (\bibinfo{year}{2010}).

\bibitem[{\citenamefont{Agrawal and Dutta}(1993)}]{agrawal1993infrared}
\bibinfo{author}{\bibfnamefont{G.~P.} \bibnamefont{Agrawal}} \bibnamefont{and}
  \bibinfo{author}{\bibfnamefont{N.~K.} \bibnamefont{Dutta}},
  \emph{\bibinfo{title}{Infrared and Visible Semiconductor Lasers}}
  (\bibinfo{publisher}{Springer}, \bibinfo{year}{1993}).

\bibitem[{\citenamefont{Barland et~al.}(2002)\citenamefont{Barland, Tredicce,
  Brambilla, Lugiato, Balle, Giudici, Maggipinto, Spinelli, Tissoni,
  Kn{\"{o}}del et~al.}}]{solitette}
\bibinfo{author}{\bibfnamefont{S.}~\bibnamefont{Barland}},
  \bibinfo{author}{\bibfnamefont{J.}~\bibnamefont{Tredicce}},
  \bibinfo{author}{\bibfnamefont{M.}~\bibnamefont{Brambilla}},
  \bibinfo{author}{\bibfnamefont{L.~A.} \bibnamefont{Lugiato}},
  \bibinfo{author}{\bibfnamefont{S.}~\bibnamefont{Balle}},
  \bibinfo{author}{\bibfnamefont{M.}~\bibnamefont{Giudici}},
  \bibinfo{author}{\bibfnamefont{T.}~\bibnamefont{Maggipinto}},
  \bibinfo{author}{\bibfnamefont{L.}~\bibnamefont{Spinelli}},
  \bibinfo{author}{\bibfnamefont{G.}~\bibnamefont{Tissoni}},
  \bibinfo{author}{\bibfnamefont{T.}~\bibnamefont{Kn{\"{o}}del}},
  \bibnamefont{et~al.}, \bibinfo{journal}{Nature}
  \textbf{\bibinfo{volume}{419}}, \bibinfo{pages}{699} (\bibinfo{year}{2002}).

\bibitem[{\citenamefont{Leo et~al.}(2010)\citenamefont{Leo, Coen, Kockaert,
  Gorza, Emplit, and Haelterman}}]{temporalcs}
\bibinfo{author}{\bibfnamefont{F.}~\bibnamefont{Leo}},
  \bibinfo{author}{\bibfnamefont{S.}~\bibnamefont{Coen}},
  \bibinfo{author}{\bibfnamefont{P.}~\bibnamefont{Kockaert}},
  \bibinfo{author}{\bibfnamefont{S.~P.} \bibnamefont{Gorza}},
  \bibinfo{author}{\bibfnamefont{P.}~\bibnamefont{Emplit}}, \bibnamefont{and}
  \bibinfo{author}{\bibfnamefont{M.}~\bibnamefont{Haelterman}},
  \bibinfo{journal}{Nature Photonics} \textbf{\bibinfo{volume}{4}},
  \bibinfo{pages}{471} (\bibinfo{year}{2010}), ISSN \bibinfo{issn}{1749-4885}.

\bibitem[{\citenamefont{Garbin et~al.}(2015)\citenamefont{Garbin, Javaloyes,
  Tissoni, and Barland}}]{garbin2015topological}
\bibinfo{author}{\bibfnamefont{B.}~\bibnamefont{Garbin}},
  \bibinfo{author}{\bibfnamefont{J.}~\bibnamefont{Javaloyes}},
  \bibinfo{author}{\bibfnamefont{G.}~\bibnamefont{Tissoni}}, \bibnamefont{and}
  \bibinfo{author}{\bibfnamefont{S.}~\bibnamefont{Barland}},
  \bibinfo{journal}{Nature communications} \textbf{\bibinfo{volume}{6}}
  (\bibinfo{year}{2015}).

\bibitem[{\citenamefont{van Saarloos}(2003)}]{van2003front}
\bibinfo{author}{\bibfnamefont{W.}~\bibnamefont{van Saarloos}},
  \bibinfo{journal}{Physics reports} \textbf{\bibinfo{volume}{386}},
  \bibinfo{pages}{29} (\bibinfo{year}{2003}).

\bibitem[{\citenamefont{Jang et~al.}(2013)\citenamefont{Jang, Erkintalo,
  Murdoch, and Coen}}]{jang2013ultraweak}
\bibinfo{author}{\bibfnamefont{J.~K.} \bibnamefont{Jang}},
  \bibinfo{author}{\bibfnamefont{M.}~\bibnamefont{Erkintalo}},
  \bibinfo{author}{\bibfnamefont{S.~G.} \bibnamefont{Murdoch}},
  \bibnamefont{and} \bibinfo{author}{\bibfnamefont{S.}~\bibnamefont{Coen}},
  \bibinfo{journal}{Nature Photonics} \textbf{\bibinfo{volume}{7}},
  \bibinfo{pages}{657} (\bibinfo{year}{2013}).

\bibitem[{\citenamefont{Erkintalo et~al.}(2015)\citenamefont{Erkintalo, Luo,
  Jang, Coen, and Murdoch}}]{erkintalo2015bunching}
\bibinfo{author}{\bibfnamefont{M.}~\bibnamefont{Erkintalo}},
  \bibinfo{author}{\bibfnamefont{K.}~\bibnamefont{Luo}},
  \bibinfo{author}{\bibfnamefont{J.~K.} \bibnamefont{Jang}},
  \bibinfo{author}{\bibfnamefont{S.}~\bibnamefont{Coen}}, \bibnamefont{and}
  \bibinfo{author}{\bibfnamefont{S.~G.} \bibnamefont{Murdoch}},
  \bibinfo{journal}{New Journal of Physics} \textbf{\bibinfo{volume}{17}},
  \bibinfo{pages}{115009} (\bibinfo{year}{2015}).

\bibitem[{\citenamefont{Gomila et~al.}(2015)\citenamefont{Gomila, Colet, and
  Walgraef}}]{gomila2015theory}
\bibinfo{author}{\bibfnamefont{D.}~\bibnamefont{Gomila}},
  \bibinfo{author}{\bibfnamefont{P.}~\bibnamefont{Colet}}, \bibnamefont{and}
  \bibinfo{author}{\bibfnamefont{D.}~\bibnamefont{Walgraef}},
  \bibinfo{journal}{Physical review letters} \textbf{\bibinfo{volume}{114}},
  \bibinfo{pages}{084101} (\bibinfo{year}{2015}).

\bibitem[{\citenamefont{Vladimirov and Turaev}(2005)}]{vladimirov2005model}
\bibinfo{author}{\bibfnamefont{A.~G.} \bibnamefont{Vladimirov}}
  \bibnamefont{and} \bibinfo{author}{\bibfnamefont{D.}~\bibnamefont{Turaev}},
  \bibinfo{journal}{Physical Review A} \textbf{\bibinfo{volume}{72}},
  \bibinfo{pages}{033808} (\bibinfo{year}{2005}).

\bibitem[{\citenamefont{Vladimirov et~al.}(2014)\citenamefont{Vladimirov,
  Pimenov, and Bandelow}}]{vladimirov2014modeling}
\bibinfo{author}{\bibfnamefont{A.~G.} \bibnamefont{Vladimirov}},
  \bibinfo{author}{\bibfnamefont{A.}~\bibnamefont{Pimenov}}, \bibnamefont{and}
  \bibinfo{author}{\bibfnamefont{U.}~\bibnamefont{Bandelow}}, in
  \emph{\bibinfo{booktitle}{Numerical Simulation of Optoelectronic Devices
  (NUSOD), 2014 14th International Conference on}} (\bibinfo{year}{2014}), pp.
  \bibinfo{pages}{153--154}.

\end{thebibliography}

\end{document}